\documentclass[acmsmall,screen,nonacm]{acmart}
\usepackage{subcaption}

\AtBeginDocument{%
  }

\acmJournal{CSUR}

\begin{document}

\title{Systematic Literature Review of Automation and Artificial Intelligence in Usability Issue Detection}
\thanks{This work was supported by the Slovak Research and Development Agency under the Contracts no. APVV-23-0408, also by the Cultural and Educational Grant Agency of Slovak Republic (KEGA) under grant no. KG 014STU-4/2024. We would like to thank UXtweak j.s.a. for their generous financial support of this research and for the technical and expert support provided by the UXtweak Research team.}

\author{Eduard Kuric}
\email{eduard.kuric@stuba.sk}
\orcid{0000-0002-7371-5512}
\affiliation{%
  \institution{Faculty of Informatics and Information Technologies, Slovak University of Technology}
  \city{Bratislava}
  \country{Slovakia}
}
\affiliation{%
  \institution{UXtweak Research}
  \city{Bratislava}
  \country{Slovakia}
}

\author{Peter Demcak}
\orcid{0000-0002-4111-1052}
\affiliation{%
  \institution{UXtweak Research}
  \city{Bratislava}
  \country{Slovakia}
}

\author{Matus Krajcovic}
\orcid{0000-0001-9030-7337}
\affiliation{%
  \institution{UXtweak Research}
  \city{Bratislava}
  \country{Slovakia}
}
\affiliation{%
  \institution{Faculty of Informatics and Information Technologies, Slovak University of Technology}
  \city{Bratislava}
  \country{Slovakia}
}

\author{Jan Lang}
\orcid{0000-0002-3271-7271}
\affiliation{%
  \institution{Faculty of Informatics and Information Technologies, Slovak University of Technology}
  \city{Bratislava}
  \country{Slovakia}
}

\renewcommand{\shortauthors}{Kuric et al.}

\begin{abstract}
Usability issues can hinder the effective use of software. Therefore, various techniques are deployed to diagnose and mitigate them. However, these techniques are costly and time-consuming, particularly in iterative design and development. A substantial body of research indicates that automation and artificial intelligence can enhance the process of obtaining usability insights. In our systematic review of 155 publications, we offer a comprehensive overview of the current state of the art for automated usability issue detection. We analyze trends, paradigms, and the technical context in which they are applied. Finally, we discuss the implications and potential directions for future research.
\end{abstract}



\maketitle

\section{Introduction}

Usability is a crucial characteristic of software that determines how well it can be used by its intended users in the appropriate context \cite{iso2018}. In the field of human-computer interaction, a variety of techniques has been designed and practiced over the years to assess usability \cite{kuric2024a, tapia2022, wang2019, piirisild2024, salvador2014}. Generally, they can be divided into
\begin{itemize}
    \item formative approaches, aimed at gaining an understanding of usability to improve it iteratively, and 
    \item summative approaches, aimed at evaluating usability by measuring constructs to ensure that systems meet usability criteria and standards \cite{lewis2014}. 
\end{itemize}

Despite differences between paradigms, they share a common high-level goal: detecting the presence of usability issues that can hinder the user experience (UX) of digital products and services. Usability research tools such as UXtweak\footnote{UXtweak usability research tool: https://www.uxtweak.com/}, Crazy Egg\footnote{Crazy Egg: https://www.crazyegg.com/} and Clarity\footnote{Microsoft Clarity: https://clarity.microsoft.com/} exist to facilitate processing of feedback that is gathered directly from users for this purpose.

However, activities aimed at manually identifying usability issues require a significant investment of time and effort. They can also be challenging to scale in the wild \cite{kuric2025}. Therefore, automation and semi-automation is a prominent area of study to assist with the collection and analysis of usability information (e.g., user behavior in systems and prototypes, video and audio recordings, eye tracking, models and images of user interfaces). Technological breakthroughs in Machine Learning (ML) and Artificial Intelligence (AI) such as more advanced Convolutional Neural Networks (CNNs) and Large Language Models (LLMs) have introduced novel ways to process data. As researchers race to explore new technologies for automatically detecting usability issues \cite{turbeville2024, kuric2024, fan2020, xing2021, esposito2022}, it is becoming increasingly challenging to understand the broader context, assess the significance of individual approaches within it, track trends and contribute to the discourse.

In this article, we address the research gap in the lack of a systematic literature review on automatic detection of usability issues. Our survey aims to provide a comprehensive and up-to-date overview, along with a thematic synthesis of the current state of the art, involving both AI and more traditional automation methods. Some secondary studies with thematic overlap have been performed in the past (see \autoref{tab:related-studies} for comparison with our survey). While valuable, they either focus primarily on separate distinct aspects (e.g., eye tracking, measuring usability as a metric) \cite{novak2023, stige2023, castro2022, abbas2022}, are not systematic literature reviews \cite{tsui2023, abuaddous2022, bakaev2016}, or provide a modest coverage of existing research \cite{stige2023, pettersson2018, abbas2022}.

\begin{table}[!ht]
\caption{Comparison of this work to the most related systematic literature reviews (SLR) and other surveys. No previous secondary research has systematically and comprehensively investigated the current state of automation and AI in usability issue detection.}
\label{tab:related-studies}
\resizebox{\columnwidth}{!}{%
\begin{tabular}{llp{4cm}llp{7cm}}
\toprule
\textbf{Ref.} & \textbf{Year} & \textbf{Focus} & \textbf{SLR} & \textbf{Corpus} & \textbf{Limitations}\\ 
\midrule
Ours & 2025 & Automated usability issue detection & \textbf{Yes} & \textbf{155} & - \\
\cite{novak2023} & 2024 & Eye tracking and its evaluation with ML & Yes & 90 & Aside from eye tracking, other usability assessment approaches were not investigated. ML automation—present only in a portion of publications—was assessed in tasks that do not directly address usability issue detection (e.g., segmentation, classification of users). \\ 
\cite{stige2023} & 2024 & Use of AI in UX processes & Yes & 46 & AI was investigated in other contexts, such as creating solutions, prototypes, specifications and personas. Non-AI automation was not discussed. Limited AI search terms (e.g., no neural networks, LLMs).\\
\cite{tsui2023} & 2023 & Automated UX evaluation in the context of traditional UX evaluation & No & - & Lacking a detailed analysis of automated approaches. As a survey that is not a systematic literature review, generalizability and robustness of findings could be questioned. \\
\cite{abbas2022} & 2022 & ML applications in UX - algorithms, techniques & Yes & 18 & Investigation primarily about challenges of applying ML in UX design contexts (e.g., design tools for ML solutions). No detailed analysis of automated usability issue detection.\\
\cite{abuaddous2022} & 2022 & Mobile automated usability evaluation & No & 19 & Survey of limited scope. Not a systematic literature review, thus raising concerns about thoroughness. Narrow specialization on mobile that did not discuss other environments (e.g., desktop, VR, wearables).\\
\cite{castro2022} & 2022 & Online automated tools to evaluate usability & Yes & 15 & Scope limited to readily available tools for quantitative evaluation only. Low-maturity methods and concepts for usability assessment were not addressed.\\
\cite{pettersson2018} & 2018 & Combination of methods for UX evaluation & Yes & 100 & Manual (non-automated) methods for evaluating dimensions of UX were the focus. Automation was discussed only briefly as an emergent technology.\\
\cite{bakaev2016} & 2016 & Methods for automated evaluation of the usability of websites & No & - & Older survey where approaches for automated usability issue detection were not thoroughly discussed. The literature review is not systematic. Narrow focus on websites, website analytics and automated collection of data.\\
\bottomrule
\end{tabular}%
}
\end{table}

The contribution of this work is focused on systematically reviewing and synthesizing findings from a broad range of primary research literature, in order to thoroughly answer questions about the context in which usability issues were detected automatically. We investigate the types of challenges faced in current research to uncover problems relevant in assessment of usability. We provide a thorough overview of which technologies, devices and types of data were incorporated and in what manner. Additionally, we also survey the trends in the use of automation and AI, then analyze their state of readiness for practical application.

The structure of this article is as follows. Section \ref{sec:rqs} introduces the research questions that we explore in our review. Section \ref{sec:methodology} establishes our methodology, describing the protocol by which the reviewed primary studies were collected and processed, alongside the details of the protocol’s execution. Section \ref{sec:results} presents our findings, followed by Section \ref{sec:discussion} which discusses observed patterns and implications in further depth, including a critical perspective. Section \ref{sec:threats} addresses the threats to the validity of our review. Finally, concluding statements are presented in Section \ref{sec:conclusion}.

\section{Research questions}
\label{sec:rqs}

Motivated by the goal of investigating the current state of knowledge about automated methods applicable for detection of usability issues, we formulated a list of research questions (see \autoref{tab:rqs}). By asking questions from a multitude of perspectives, we seek to analyze primary studies generally yet comprehensively, with regard to the addressed problems, common approaches, sources of data and current trends. Research questions support the planning of the literature review protocol.

\begin{table}[!ht]
\caption{Research questions addressed by this systematic literature review.}
\label{tab:rqs}
\resizebox{\columnwidth}{!}{%
\begin{tabular}{llp{13cm}}
\toprule
\textbf{\#} & \textbf{Title} & \textbf{Question} \\
\midrule
RQ1 & Chronology & What is the temporal distribution of research dedicated to automated usability issue detection by publication year? \\
RQ2 & Objectives & In the context of what specific research objectives is automated usability issue detection used? \\
RQ3 & Technology & What intelligent technologies have been assessed for automated usability issue detection? \\
RQ4 & Artificial intelligence & To what degree is artificial intelligence applied to automated usability issue detection? \\
RQ5 & Devices & On which types of computing devices has automated usability issue detection been examined? \\
RQ6 & Data & What data sources are employed for automated usability issue detection? \\
RQ7 & Maturity & What is the technological maturity of automated usability issue detection approaches? (e.g., concept, prototype, evaluated instrument) \\
RQ8 & Participant involvement & What is the involvement of participants in automated usability issue detection? \\
\bottomrule
\end{tabular}%
}
\end{table}

RQ1 to RQ6 are used to assess general descriptive factors of the state of the art, to investigate when, why and how have studies of usability issue automation been conducted. Because of the rapid development of machine learning and artificial intelligence (AI) in recent years \cite{collins2021, neu2022}, RQ3 explores the underlying technologies and RQ4 further analyzes the representation of AI. RQ7 explores the readiness of investigated approaches, from concept to tools available for real-world use. With RQ8, we address the concern of whether automated usability findings are based on information from genuine participants (users) in order to capture dimensions of usability in realistic use contexts \cite{namoun2021}.

\section{Methodology}
\label{sec:methodology}

\begin{figure}[!ht]
  \centering
  \includegraphics[width=\linewidth]{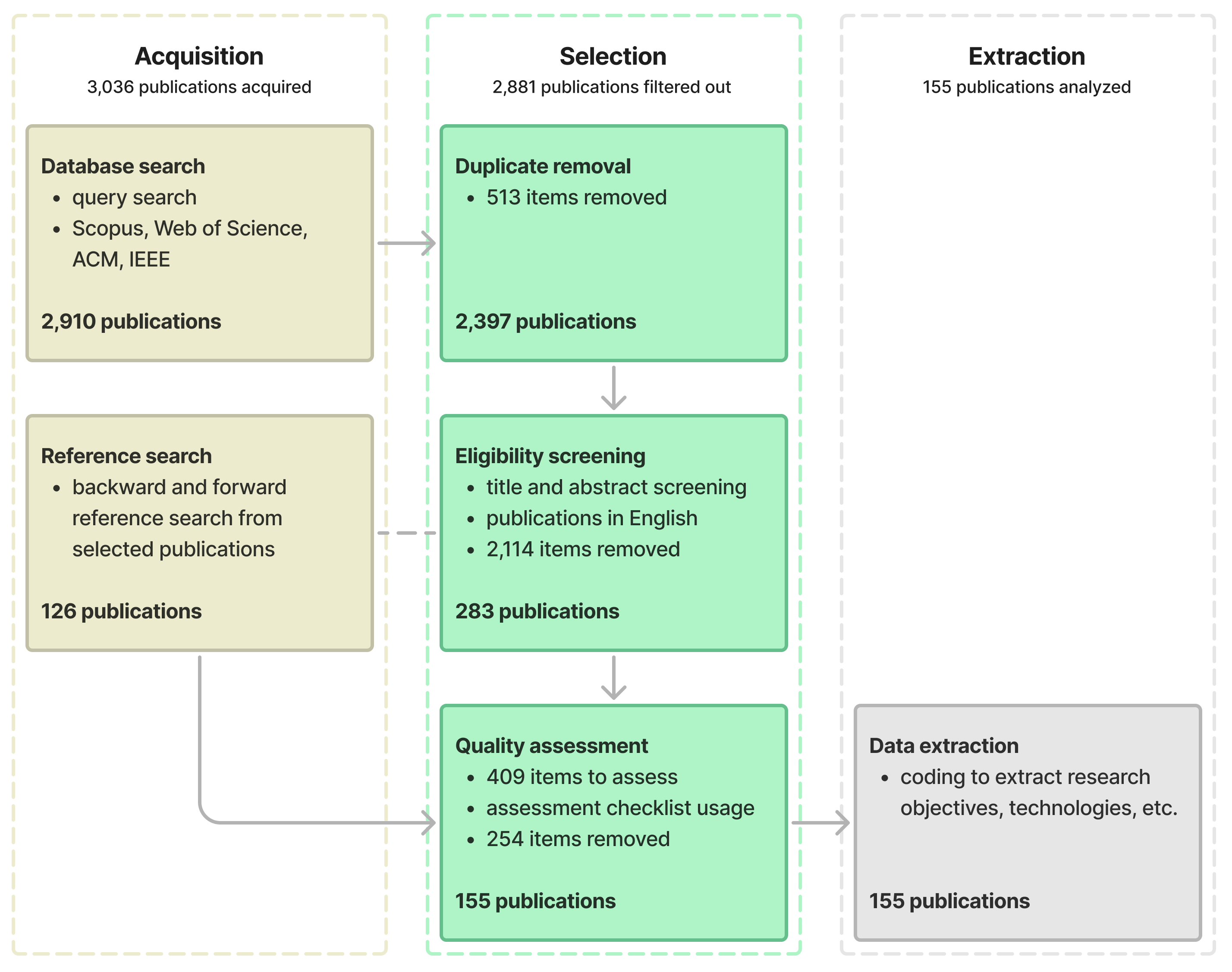}
  \caption{Funnel diagram of the systematic literature survey protocol, portraying the downward filter of publications to obtain unique, relevant and high-quality articles. The acquisition, selection, and extraction process resulted in 155 publications being included for analysis.}
  \Description{The image is a flowchart illustrating the execution of systematic literature survey protocol. It is composed of three main stages: Acquisition, Selection, and Extraction. The Acquisition phase involves database and reference searches, in which 2,910 publications and 126 publications were acquired respectively. Selection includes steps such as duplicate removal, eligibility screening, and quality assessment, leaving 155 publications eligible for further analysis. The process concludes with data extraction.}
  \label{fig:diagram}
\end{figure}

Our systematic literature review meticulously adheres to the guidelines devised for secondary studies in computer science by \citet{carrera-rivera2022}. Their guidelines provide our protocol template with clear steps (see \autoref{fig:diagram}). First, digital library sources were selected and inclusion and exclusion conditions were determined through the process of gradual refinement. Deduplicated articles from the search results were first screened by their titles and abstracts, followed by a fine-grained analysis of their contents. To broaden the scope of the survey, relevant citations and references of the search results were also included in the article pool.

To resolve researcher bias and achieve higher reliability of our literature review, the decision-making process involved the collaboration of three researchers. Articles were concurrently screened and classified by the agreement of two assessors. A third assessor resolved any disagreements that emerged.

\subsection{Database selection}

Comprehensive analysis of the state of the art demands that we cover reliable publications from highly-reputable sources. The intersection between usability and artificial intelligence is an interdisciplinary subject. Therefore, for our investigation, we selected the world’s largest multidisciplinary databases (Scopus\footnote{Scopus: \url{https://www.scopus.com/}}, Web of Science\footnote{Web of Science: \url{https://www.webofscience.com/}}), as well as databases specific to engineering and technology (IEEE Xplore Digital Library\footnote{IEEE XPlore Digital Library: \url{http://ieeexplore.ieee.org/}}) and computer science (ACM Digital Library\footnote{ACM Digital Library: \url{https://dl.acm.org/}}). These databases offer advanced search query capabilities and are frequently cited as sources in surveys and literature reviews within related fields \cite{carrera-rivera2022, zuo2023, abbas2022, kong2019, quin2024}.  

\subsection{Query search method and duplicate removal}

For the search query to include terms that are pivotal to the topic, including common related words and synonyms, search keywords were developed through iterative and collaborative exploration. Three categories of keywords were incorporated to increase the flexibility of the filter. Across their title, keywords and abstract, the articles must contain at least one word from each following category:

\begin{enumerate}
    \item usability, ux, user experience
    \item test*, evaluat*, issue*, problem*, smell*, research
    \item automat*, ai, intelligen*, artificial intelligence, chatgpt, llm, ml, machine learning, nn, neural network, deep learning, gpt
\end{enumerate}

Wildcards (*) were used in the queries to account for variability in suffixes, such as gerunds for verbs and plurals for nouns. For instance, test* will also match words like testing, tested, and tests. In the queries, proximity operators were used instead of simple boolean operators to obtain more relevant results (precision was boosted from 2\% to 13\%). Due to differences in search functionality available across digital libraries, the queries engineered to retrieve publications vary slightly, as seen in \autoref{tab:query-strings}.

\begin{table}[!ht]
\caption{Query strings used to perform consistent search across digital libraries and the number of publications yielded by each query. Search results were acquired on 2024-12-11. ACM Digital Library does not support proximity operators, thereby necessitating the use of boolean conjunction.}
\label{tab:query-strings}
\resizebox{\columnwidth}{!}{%
\begin{tabular}{lp{13.1cm}l}
\toprule
\textbf{Source} & \textbf{Search string} & \textbf{Results} \\
\midrule
Scopus & TITLE-ABS-KEY ( ( usability OR ux OR "user experience" ) W/5 ( test* OR evaluat* OR issue* OR problem* OR smell* OR research ) W/5 ( automat* OR ai OR intelligen* OR "artificial intelligence" OR chatgpt OR llm OR ml OR "machine learning" OR nn OR "neural network" OR "deep learning" OR gpt ) ) & 1232 \\
\midrule
Web of Science & TS=( ( usability OR ux OR "user experience" ) NEAR/5 ( test* OR evaluat* OR issue* OR problem* OR smell* OR research ) NEAR/5 ( automat* OR ai OR intelligen* OR "artificial intelligence" OR chatgpt OR llm OR ml OR "machine learning" OR nn OR "neural network" OR "deep learning" OR gpt ) ) & 436 \\
\midrule
IEEE Digital Library & ( ( "All Metadata":usability OR "All Metadata":ux OR "All Metadata":"user experience" ) NEAR/5 ( "All Metadata":test* OR "All Metadata":evaluat* OR "All Metadata":issue* OR "All Metadata":problem* OR "All Metadata":smell* OR "All Metadata":research ) NEAR/5 ( "All Metadata":automat* OR "All Metadata":ai OR "All Metadata":intelligen* OR "All Metadata":"artificial intelligence" OR "All Metadata":chatgpt OR "All Metadata":llm OR "All Metadata":ml OR "All Metadata":"machine learning" OR "All Metadata":nn OR "All Metadata":"neural network" OR "deep learning" OR gpt ) ) & 770 \\
\midrule
ACM Digital Library & Title:(( ( usability OR ux OR "user experience" ) AND ( test* OR evaluat* OR issue* OR problem* OR smell* OR research ) AND ( automat* OR ai OR intelligen* OR "artificial intelligence" OR chatgpt OR llm OR ml OR "machine learning" OR nn OR "neural network" OR "deep learning" OR gpt ) )) & 472 \\
\midrule
\multicolumn{2}{l}{\textbf{\textit{SUM}}} & \textbf{2910} \\
\bottomrule
\end{tabular}%
}
\end{table}

The publication filter included both journal articles and conference papers. To review works that were current at the time of acquisition (2024-12-11), the publication date filter spanned the period of the previous ten years (2014–2024). Our analysis revealed that automated usability issue detection started to gain more traction around 2016. To keep pace with the rapidly evolving field of AI automation, preprint articles were included to undergo a separate evaluation during the quality assessment step.

In total, 2,910 publications were retrieved from database searches. The deduplication process using Zotero\footnote{Zotero research assistant software: \url{https://www.zotero.org/}} software yielded 2,397 unique articles by resolving overlaps between digital libraries that index articles independently.

\subsection{Eligibility screening}

To efficiently eliminate articles that were unrelated to automated identification of usability issues, we evaluated their titles and abstracts. The screening filter reduced the corpus to exclusively feature primary research. The exclusion criteria included irrelevance to the topic and the research questions, and a publication language other than English. In total, 283 publications passed the screening, with 2,114 publications screened out, most of them due to involving usability evaluation without any intelligent automation, applications of UX methods in AI-driven systems, machine learning or deep learning research unrelated to usability, and other topics that encompass the filter keywords in an irrelevant context.

\subsection{Reference search}

Considering disparities in terminology, keyword search does not always capture a holistic view of relevant papers. Snowball search \cite{wohlin2014, wohlin2022} is a complementary method that begins with a small starting set of typically 5–10 highly cited articles in the field and expands outwards by iteratively searching references and citations. However, snowballing can raise concerns about bias due to its dependence on the starting set selection, as well as its efficiency \cite{jalali2014}.

We applied a hybrid approach where the large set of 283 eligible results from the extensive keyword search provided the basis for a single-iteration reference and citation search. Additional 126 articles were retrieved after the results underwent an identical screening process to the keyword search. Maintaining consistency with the eligibility screening, some references with similar subjects to their citations were not included (e.g., studies involving adjacent methods, but without focus on automated usability issue detection).

\subsection{Quality assessment}

Managing the quality of publications is essential in secondary research to prevent systematic errors and perpetuation of biases \cite{kitchenham2010, dyba2008}. Instruments for quality assessment (QA) are typically checklists, some established and widely adopted while additional customization can also be introduced to align with the research needs \cite{yang2021b, ambreen2018}. We performed three checks to verify relevance, rigor and credibility. Each aspect was scored on a range from 0 to 1.

Relevance to the Research Questions (QA1) was investigated in detail to address the lack of specificity in the abstracts during eligibility screening. The score of 1 was reserved for publications concerned directly with automated usability issue detection. Publications where usability issue detection was not the primary aim, yet the results could be interpreted as topical were deemed to meet the criterion partially, receiving the score of 0.5.

Methodological and Reporting Rigor (QA2) was assessed based on the comprehensive checklist by \citet{kitchenham2010}. Studies were examined to determine whether they sufficiently describe factors including aims, research questions, experiment design, procedure, biases, analysis, findings and implications. Supplementary focus was placed on the presence of information to be retrieved during the data extraction step. Publications that provide full information clearly and explicitly received a score of 1. A score of 0.5 was assigned to publications where extraction of salient information was hindered by implicitness or lack of clarity, thus they were deemed to fulfill the criterion partially.

Source Credibility (QA3) focused on the acceptance of the primary studies by the research community, represented by the rank of the source journal or conference during the publication year. For journals, Journal Citation Report\footnote{Journal citation report: \url{https://jcr.clarivate.com/}} (JCR) rankings were assessed. In case of missing entries, Scimago Journal Rank\footnote{Scimago journal rank: \url{https://www.scimagojr.com/}} (SJR) and the nearest available year’s entry were consulted as fallback. Each journal was assigned its best quartile (Q1–Q4) in categories related to computer science (including multidisciplinary subfields linked to sociology, psychology or ergonomics). Conference rankings were retrieved from ICORE\footnote{ICORE Conference portal: \url{https://portal.core.edu.au/conf-ranks/}}, with the Conference ranks\footnote{Conference ranks: \url{http://www.conferenceranks.com/}} website serving as fallback. A score of 1 was assigned to journals ranked Q1–Q3 and conferences rated A*, A or B by ICORE or ERA, or A1, A2, B1 or B2 by Qualis. 

To account for ongoing dynamic developments in the field of automation and AI, we provided high-impact publications with an alternative path to demonstrate broad acceptance despite not yet being formally published or being published in lower-ranked venues. Publications that accumulated at least five citations on Google Scholar per year on average received a Source Credibility score of 1. A high threshold was used as a precaution against auto-citations. See \autoref{tab:qas} for a summary of quality assessment questions.

\begin{table}[!ht]
\caption{List of three quality assessment questions designed to evaluate the overall quality of publications and their eligibility for inclusion in our dataset.}
\label{tab:qas}
\resizebox{\columnwidth}{!}{%
\begin{tabular}{llp{7cm}l}
\toprule
\textbf{\#} & \textbf{Aspect} & \textbf{Question} & \textbf{Values} \\
\midrule
QA1 & Relevance to Research Questions & \textit{Does the publication examine automated approach(es) of detecting usability issues?} & Yes = 1, Partially = 0.5, No = 0 \\
QA2 & Methodological and Reporting Rigor & \textit{Does the publication holistically describe a rigorous research methodology?} & Yes = 1, Partially = 0.5, No = 0 \\
QA3 & Source Credibility & \textit{Is the publication from a credible source?} & Yes = 1, No = 0 \\
\bottomrule
\end{tabular}%
}
\end{table}

As the inclusion criterion, publications needed to achieve a score of 2.5, accommodating studies that lose 0.5 of the score either by being highly relevant but with some clarity issues, or high-quality with ancillary relevance due to having a different primary aim. In total, 155 publications passed the quality assessment and were included in further analysis.

\subsection{Data extraction}

Information relevant for answering research questions (see \autoref{tab:rqs}) was extracted from individual publications. For operational and descriptive purposes, bibliographic data extracted included the title, authors, abstract, keywords, journal or conference name, publisher name (unified as it appears in JCR), DOI, issue and volume number, number of citations and journal/conference rankings. Citation counts were extracted from Google Scholar on 2024-12-11. All obtained information is available as a spreadsheet in the public repository (see \hyperref[sec:data-statement]{Additional materials}).

The coding process was inductive. Objectives, maturity level and participant involvement were coded as single-label, while technology, data sources, and device types were multi-label.

\subsection{Data analysis}

Python was employed for both data analysis and bibliographic data preparation. Libraries that were utilized include Pandas for handling CSV data, Pybtex for managing bibliographic data, and Wordcloud for keyword analysis. Statistical analyses involved the Spearman’s correlation coefficient to assess relationships in data.

\section{Results}
\label{sec:results}

\begin{figure}[!ht]
    \captionsetup[subfigure]{skip=1em}
    \centering
    \begin{subfigure}[t]{0.4\textwidth}
        \centering
        \includegraphics[height=2in]{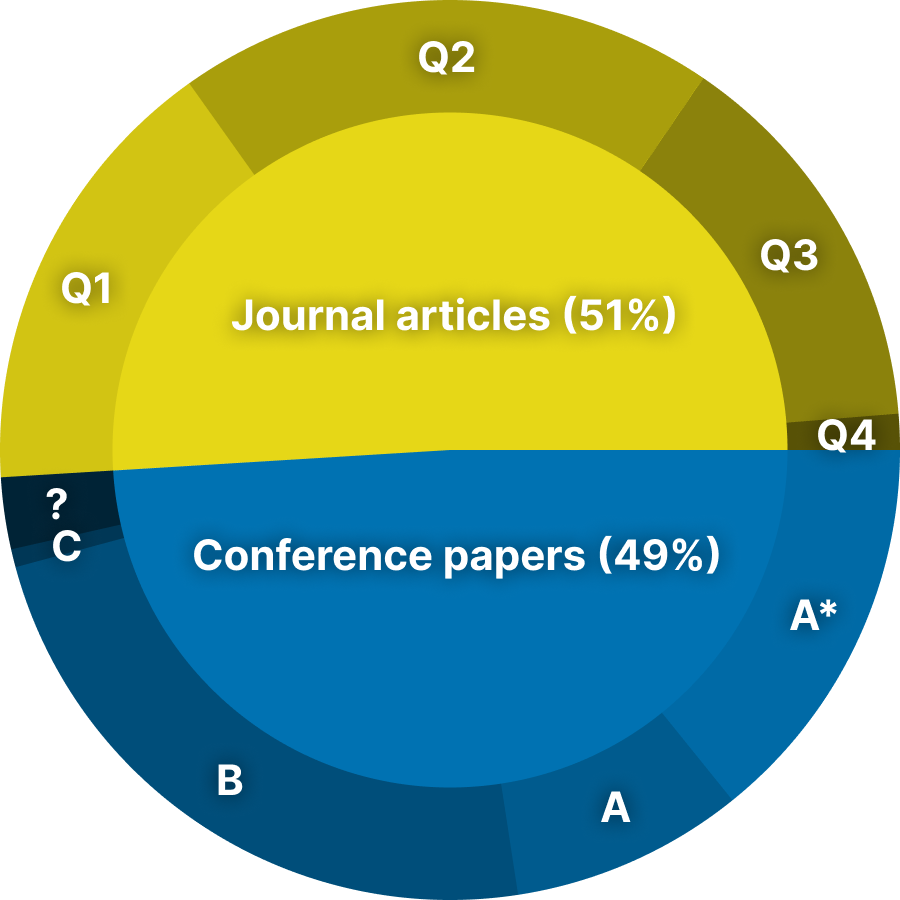}
        \caption{Publication distribution.}
        \Description{The image is a pie chart, representing the distribution of journal articles and conference papers within the dataset. Journal articles make up 51\% of all publications, while conference papers constitute 49\%. Each category is further segmented based on quartiles for journals and conference rankings for conference papers. The proportions of A and B-ranked conferences, as well as Q1, Q2, and Q3 journal articles, are similar. Only a small fraction of publications belong into the Q4 articles, or papers from C-ranked or unranked conferences.}
        \label{fig:type}
    \end{subfigure}%
    ~
    \begin{subfigure}[t]{0.6\textwidth}
        \centering
        \includegraphics[height=2.2in]{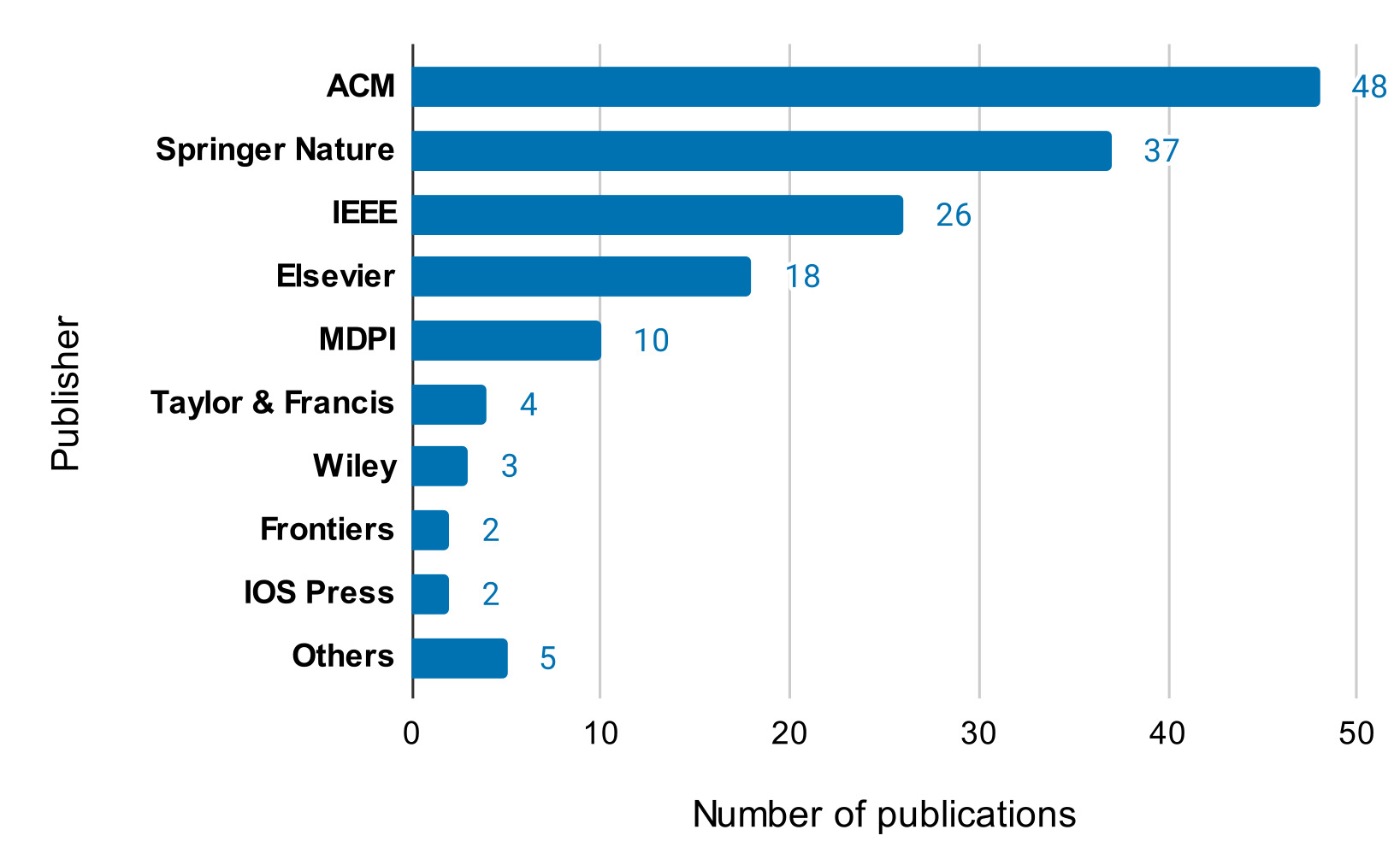}
        \caption{Publisher distribution.}
        \label{fig:publ}
        \Description{The image is a horizontal bar chart displaying the number of publications per publisher. The horizontal axis represents the number of publications, while the vertical axis lists the publishers. ACM is the most prevalent publisher, with 48 publications, followed by Springer Nature with 37, IEEE with 26 and Elsevier with 18 publications. Less common publishers in the corpus, ordered by the number of publications, include MDPI, Taylor & Francis, Wiley, Frontiers and IOS Press, each with 10 publications or less.}
    \end{subfigure}
    \caption{Distribution of primary studies in the corpus, categorized by journal impact quartiles and conference rankings (a) and primary study publisher distribution bar chart (b). The question mark (?) represents unranked conferences. The dataset shows similar ratios of journal articles and conference papers, along with their quality indicators. The most prevalent publishers include ACM with 48 publications in total, followed by Springer Nature, IEEE, and Elsevier.}
\end{figure}

This section presents the analyses and findings related to the research questions. The analyzed primary studies originate from a balanced ratio of journals and conferences of varied impact factors and ranks (see \autoref{fig:type}). The prevalent publishers include ACM, Springer Nature, IEEE, and Elsevier (see \autoref{fig:publ}). The most frequent terms among keywords include the words \textit{web}, \textit{usability}, and \textit{user interface}.

\subsection{Chronology (RQ1)}

\begin{figure}[!ht]
  \centering
  \includegraphics[width=0.6\linewidth]{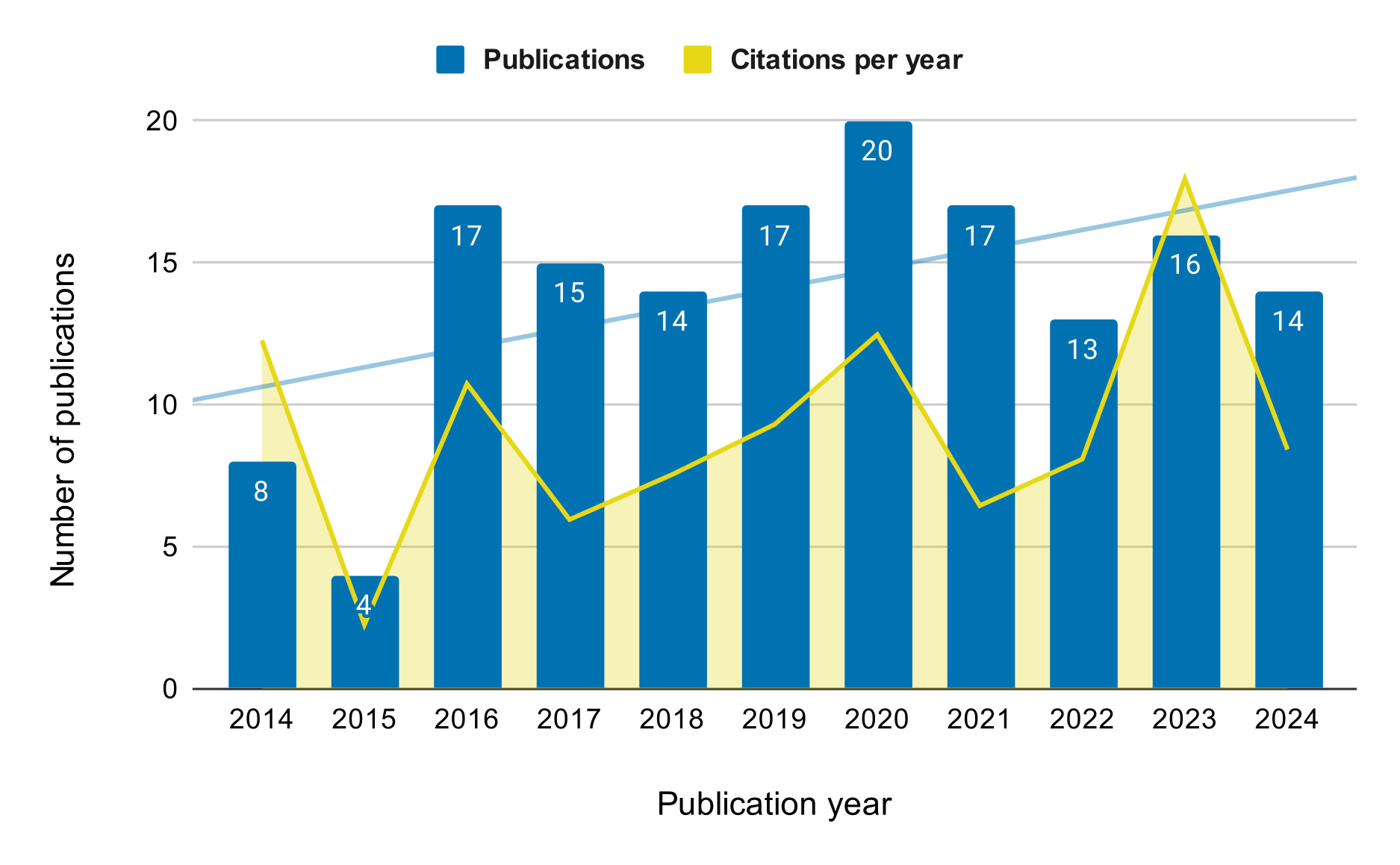}
  \caption{Yearly distribution of research publications indicates a slight upward trend, with per-year citation counts showing a notable increase over the past three years.}
  \Description{The image is a vertical bar chart displaying the number of publications and citations per year. The horizontal axis represents the publication year from 2014 to 2024 while the vertical axis represents the number of publications and citations for individual years. The number of publications and citations has grown over the years, with a minimum of 4 publications in 2015 and reaching a peak of 20 publications in 2020. The peak of per year citations is reached in 2023.}
    \label{fig:years}
\end{figure}

Between 2014 and 2024, the distribution of publications that encompass automation of usability issue detection demonstrates a slight upward trend (see \autoref{fig:years}). Similar growth can be observed in Citations per Year (see \autoref{eq:citations}), which normalizes the number of citations by the time elapsed since their publication. 

\begin{equation}
    \label{eq:citations}
    CitationsPerYear = \frac{TotalCitations}{Years Since Publication}
\end{equation}

Results suggest a growing interest in the subject. However, a period of 10 years is too short to establish statistical significance of such a slight trend (Spearman’s correlation for number of publications: $r_s(10) = .28, p = .4$; for Citations per Year: $r_s(10) = .11, p = .15$).

\subsection{Objectives (RQ2)}

\begin{figure}[!ht]
  \centering
  \includegraphics[width=0.8\linewidth]{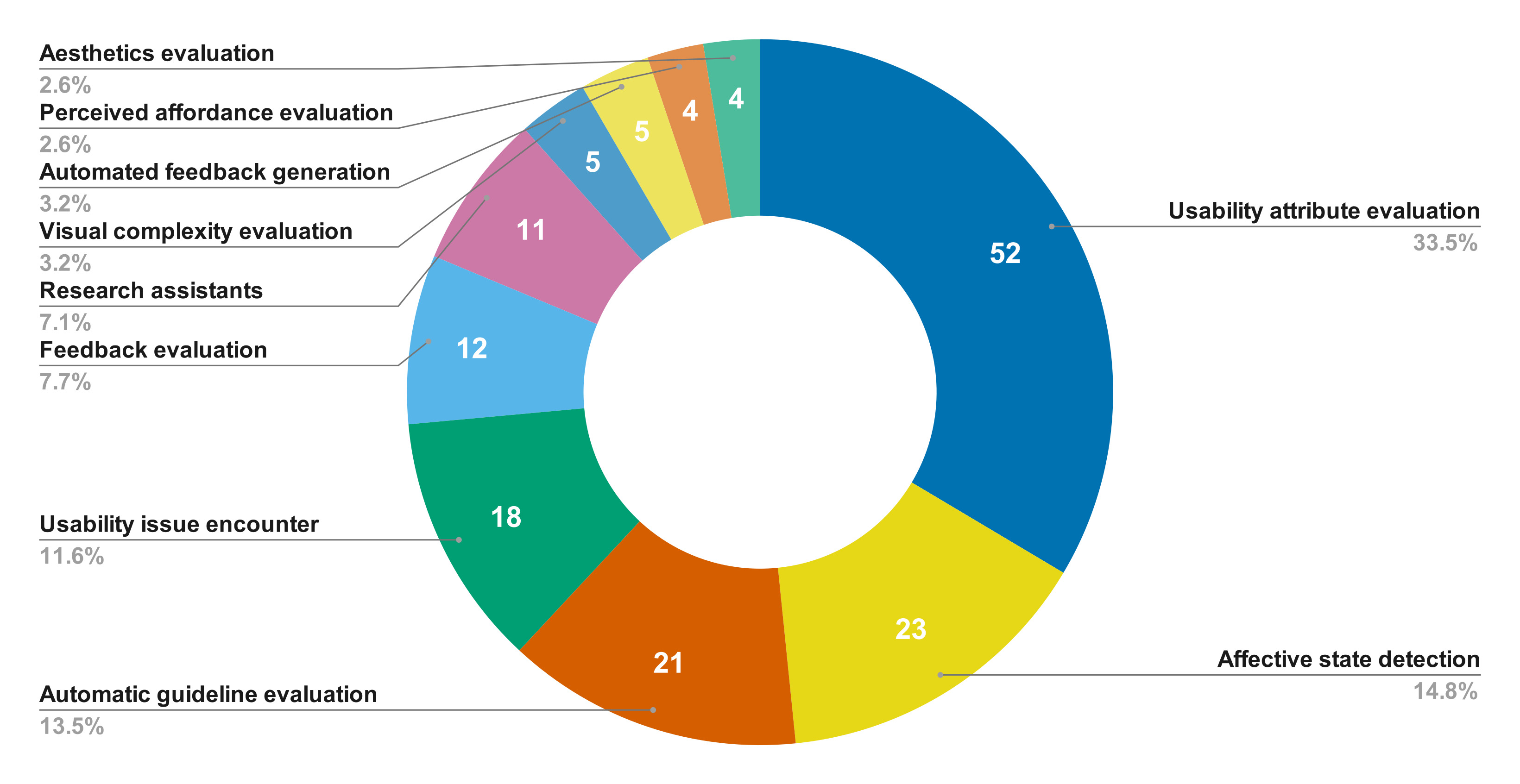}
  \caption{Distribution of research objectives in publications on automated detection of usability problems. The most common objective is Usability attribute evaluation, accounting for 34\% of publications, followed by Affective state detection and Automatic guideline evaluation.}
  \Description{The image is a pie chart displaying the proportion of research objectives for publications in the dataset. The largest segment is Usability attribute evaluation with 52 publications (or 34\%) and Affective state detection with 23 publications (or 15\%). Other objectives, in order, are Automatic guideline evaluation, Usability issue encounter detection, Feedback evaluation, Research assistants, Visual complexity evaluation, Automated feedback generation, Perceived affordance detection and Aesthetics evaluation.}
\label{fig:area-pie}
\end{figure}

Our analysis revealed that publications with implications for automated usability issue detection have diverse goals and motivations.  \autoref{fig:area-pie} shows an overview of our thematic categorization of study objectives. We identified 10 categories of objectives in total (see \autoref{tab:objectives}). Primary studies that focus directly on detecting usability issue encounters form only 11.6\% (n=18) of the sample.  Depending on the context of the usability assessment in a study, implications for usability issue detection range in their explicitness. In cases when publications could be logically placed into multiple categories due to their multifaceted contents (e.g., an assistant aimed at identifying usability issue encounters from transcripts), we favored the the more specific category for which they are more relevant. Below, we investigate the categories ordered by the number of publications that they represent.

\begin{table}[!ht]
\caption{List of research objectives identified in the publication dataset with corresponding descriptions.}
\label{tab:objectives}
\resizebox{\columnwidth}{!}{%
\begin{tabular}{lp{8cm}p{5cm}}
\toprule
\textbf{Objective} & \textbf{Description} & \textbf{Publications} \\
\midrule
Aesthetics evaluation & Automated assessment of an interface's visual appeal. & \cite{dou2019, gu2021, kyelem2023, xing2021} \\
Affective state detection & Inferring users' emotions, stress, or engagement levels. & \cite{chen2021, desolda2021, esposito2024, esposito2022, feijofilho2016, filho2015, georges2016, giroux2021, kwon2022, liapis2021, liapis2021a, razzaq2023, santos2022, soleimani2017, stefancova2018, gannouni2023, hussain2018, issa2020, le-pailleur2020, liapis2014, maier2019, matlovic2016, razzaq2020} \\
Automated feedback generation & Generating human-like feedback, such as for UI improvements. & \cite{duan2023, hämäläinen2022, xiang2024, zhou2020, hämäläinen2023} \\
Automatic guideline evaluation & Automating usability inspection using guidelines and heuristics. & \cite{bacikova2021, baek2016, cassino2015, dingli2014, eskonen2020, hasanmansur2023, liu2023, marenkov2016, marenkov2016a, marenkov2017, mathur2018, moran2018, ponce2018, robal2017, soui2020, widodo2023, yang2021, zhao2020, almeida2015, hsueh2024, meixner2021} \\
Feedback evaluation & Analyzing user reviews through sentiment analysis or topic classification. & \cite{asnawi2023, bakiu2017, eiband2019, eiband2020, khalajzadeh2023, krause2017, sanchis-font2019, sanchis-font2021, wang2020, guzman2014, maalej2016, sadigov2024} \\
Perceived affordance evaluation & Identifying key interface elements using saliency maps or similar techniques. & \cite{cheng2023, koch2016, schoop2022, xu2016} \\
Research assistants & AI-powered tools assisting UX researchers in data collection and analysis. & \cite{soure2022, celino2020, batch2024, liu2024, bisante2024, kim2019, fan2022, kuang2023a, kuric2024, kuang2024, turbeville2024} \\
Usability attribute evaluation & Measuring usability aspects like efficiency, effectiveness, and satisfaction. & \cite{quade2014, kang2014, wu2015, arapakis2016, frey2016, kiseleva2016, ferre2017, aleksander2018, trinh2018, amrehn2019, buono2019, blanco-gonzalo2014, dahri2019, harrati2016, choi2019, speicher2014, poore2017, souza2022, katerina2018, cruzgardey2020, duan2020, buono2020, bures2020, yang2020, villamane2024, yu2018, koonsanit2021, kaur2016, kaur2017, prezenski2017, ismailova2017, ismailova2017a, shojaeizadeh2019, zarish2019, holzwarth2021, souza2019, parajuli2020, li2022, paul2020, verkijika2018, gardey2022, asemi2022, salomon2023, al-sakran2021, namoun2021, csontos2021, macakoglu2023, kumar2023, gupta2023, sola2024, gardey2024, fakrudeen2024} \\
Usability issue encounter detection & Detecting moments where users face difficulties in a user interface. & \cite{fan2020, fan2020a, goncalves2016, grigera2014, grigera2017a, grigera2017, harms2019, jeong2020, paterno2016, paterno2017, ribeiro2019, tapia2022, vigo2017, benvenuti2021, fan2019, firmenich2019, jorritsma2016, kaminska2022} \\
Visual complexity evaluation & Automatically assessing how visually complex or simple a design appears. & \cite{bakaev2018, bakaev2018a, boychuk2019, miniukovich2018, oulasvirta2018} \\
\bottomrule
\end{tabular}%
}
\end{table}

\begin{figure}[!ht]
  \centering
   \includegraphics[width=0.7\linewidth]{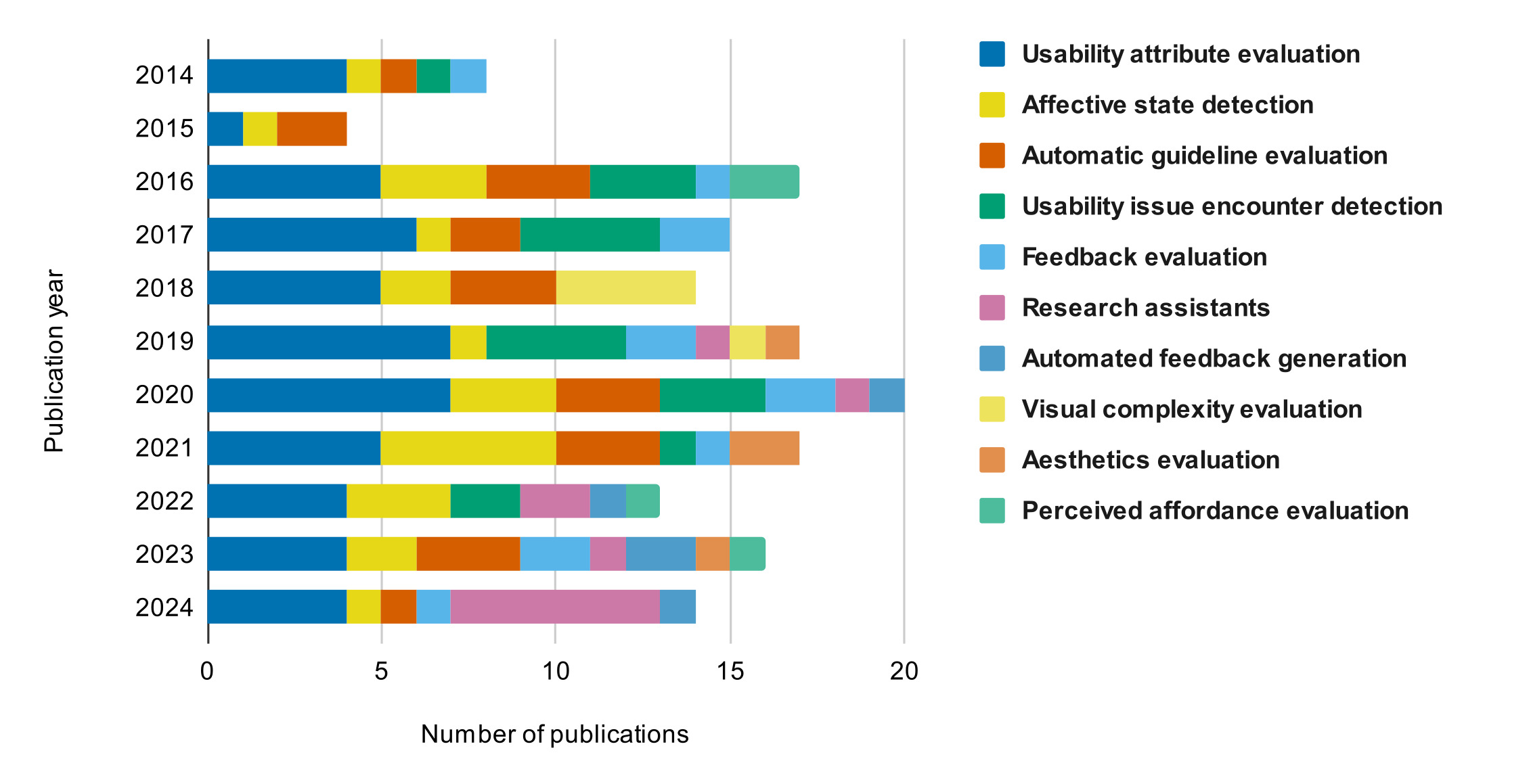}
  \caption{Per-year distribution of research objectives involving automated usability problem detection. Research assistants have become more prevalent in recent years.}
  \Description{The image is a horizontal stacked bar chart displaying the distributions of research objectives across publications for each year. The horizontal axis represents the number of publications while the vertical axis represents the publication year from 2014 to 2024. Distributions of research objectives are displayed as differently colored segments in each bar. Distributions are similar across years. Feedback generation and research assistants are gaining popularity in recent years, while usability issue encounter detection is on the decline.}
  \label{fig:area-years}
\end{figure}

On the annual basis, the distribution of research objectives has been mostly consistent between 2014 and 2024 (see \autoref{fig:area-years}). More recently, the burgeoning proliferation of AI and LLMs could be seen as the catalyst for the emergence of two novel topics of research: feedback generation and research assistants. Emotion detection signified its peak in 2020 but has been on the decline in frequency since.

\subsubsection{Usability attribute evaluation}

Encompassing the quantitative and summative evaluation of one or multiple constructs, usability attribute evaluation is the most prevalent category in the literature at 33.5\% (n=52). Attributes can serve as indicators for automatic detection of usability issues in a system. By their nature as constructs, the attributes can be further subdivided into a variety of categories as illustrated in the remainder of this section. For a brief summary, the measured attributes include models of usability and its aspects (e.g., effectiveness, learnability) and measures of human-computer interaction (e.g., effort, task difficulty).

Standards, such as the ISO 9241-11, provide robustly defined components of usability, which  \citet{dahri2019} evaluated based on user interaction. \citet{villamane2024} utilized this approach in a methodology to automate usability testing. \citet{salomon2023} instead drew from Quality-in-Use attributes defined in the ISO/IEC 25010—efficiency, effectiveness, satisfaction, freedom from risk and context coverage. \citet{asemi2022} combined metrics from standards and literature for speech-based evaluation.

Standardized user-reported usability questionnaires such as the System Usability Scale (SUS) or AttrakDiff were researched for automation using several approaches. For automation, \citet{amrehn2019} predicted an approximation of responses based on interaction data while \citet{harrati2016} explored the relationship of SUS scores and interaction metrics, demonstrating their complementary nature for assessing user acceptance. \citet{souza2019} leveraged mouse tracking, fuzzy logic and clustering for a method yielding results comparable to SUS. \citet{souza2022} applied eye tracking alongside mouse and keyboard interactions for machine learning prediction of whether the user is experienced or inexperienced based on the SUS. However, SUS is traditionally used for assessment of system usability rather than the subject’s experience level.  In VR, head yaw can correlate with SUS scores, workload scales of the NASA TLX and presence \cite{holzwarth2021}.

As an alternative to standard usability measuring instruments, heuristics-based approaches have incorporated their inherent conceptual models. \citet{speicher2014} introduced Usability-based Split Testing as a methodology that leverages interaction data for machine-learning-based prediction of informativeness, understandability, confusion, distraction, readability, information density and accessibility. \citet{li2022} proposed a rule-based framework for assessing compliance with usability requirements, including efficiency, effectiveness, learnability, operability, visibility and fault tolerance.

Specific attributes linked to measurement of usability in literature can be categorized as either system-based or interaction-based. The system-based attributes are a significant subject of investigation typically with tools that analyze source code data of the system rather than user’s interaction with it \cite{al-sakran2021, namoun2021, parajuli2020, paul2020, kaur2016, macakoglu2023}. They include page speed \cite{kaur2016, al-sakran2021, macakoglu2023, csontos2021, ismailova2017, ismailova2017a, kaur2017}, successes \cite{al-sakran2021}, efficiency \cite{namoun2021}, navigation \cite{namoun2021, verkijika2018}, information architecture \cite{namoun2021}, errors \cite{al-sakran2021, parajuli2020, csontos2021}, warnings \cite{al-sakran2021, parajuli2020}, download times \cite{al-sakran2021, paul2020} or page sizes \cite{al-sakran2021, paul2020, kaur2016, csontos2021, ismailova2017, ismailova2017a, kaur2017}, readability \cite{kaur2017} or overall performance \cite{zarish2019}. Source code data has also been applied to predict efficiency and learnability on mobile devices \cite{prezenski2017} by crawling the app and using cognitive user models.

Interaction-based usability attributes such as effort, time on task, engagement and satisfaction are the outcomes of interaction rather than the system’s intrinsic properties. \citet{cruzgardey2020} estimated effort in automated A/B testing. \citet{gardey2022} predicted interaction effort from mouse dynamics in web interface widgets. \citet{quade2014} applied deep learning to the time on task from models of user interfaces. Engagement can be predicted from interaction data \cite{kumar2023, arapakis2016}, or physiological signals mapped to multimedia context \cite{poore2017}, or as long-term engagement with agent-based intervention \cite{trinh2018}. Satisfaction was predicted for conversational assistants while incorporating characteristic control mechanisms such as voice inputs and touch gestures \cite{kiseleva2016, choi2019}. Accounting for changes in satisfaction over time, \citet{koonsanit2021} demonstrated potential for predicting the final satisfaction based on momentary satisfaction.

Certain approaches for usability attribute evaluation rely on specialized sensors. For usability evaluation, electrical brain activity (EEG) was used to infer user experience \cite{aleksander2018} and mental workload in VR \cite{frey2016}. Eye tracking data was used in machine learning by \citet{shojaeizadeh2019} to predict task load  based on eye movements. \citet{yu2018} predicted the UX of mobile games based on physiological signals like heart rate, blood pressure and oxygen saturation.

Summative evaluation through synthetic participants that simulate  user personas is an objective thematically related to feedback generation. \citet{gupta2023} explored the evaluation of consistency, efficiency, satisfaction, learnability and memorability with AI on a conceptual level.

In adaptive user interfaces and prototypes, predictions have further been leveraged to optimize their design. \citet{duan2020} employed long short-term memory networks to predict task performance and improve UI layout. \citet{kang2014} explored action sequence mining as a means for evaluating UIs in games.

\subsubsection{Affective state detection}

Research that leverages methods of affective computing for detection of usability issues comprises 14.8\% (n=23) of the corpus. It typically focuses on recognition of emotions (such as Ekman’s universal emotions: enjoyment, sadness, fear, anger, disgust, contempt and surprise) or psychophysiological state, like stress. 

A prominent emotion-driven approach for detecting usability issues involves generating emotion visualizations (typically heatmaps) based on facial expressions and interactions during usability testing \cite{esposito2024, esposito2022, filho2015} and in the wild \cite{desolda2021}. For evaluation on mobile devices, \citet{feijofilho2016} also incorporated contextual data such as location and weather. \citet{georges2016} combined gaze tracking with physiological and behavioral signals to map emotional state to targets of the user’s visual attention. 

\citet{giroux2021} proposed guidelines for automatically collecting facial data during remote unmoderated usability testing. For games, \citet{kwon2022} proposed a framework for data collection and deep learning analysis of facial expressions and engagement to evaluate game experience, potentially enhancing observation in real time. For applications that involve full-body gestures and body language, \citet{razzaq2020} predicted emotions with skeletal joint features.

Speech-based emotion detection figures prominently in usability evaluation involving the think-aloud protocol and speech-based user interfaces like conversational assistants. Emotions expressed while thinking aloud can be analyzed with machine learning to assist UX evaluation \cite{soleimani2017}. Speech sequences can be analyzed with neural networks to improve user interfaces and experiences \cite{chen2021, issa2020}. Additionally, \citet{razzaq2023} presented a multi-modal deep learning framework that predicts emotions by integrating speech, body language and facial expressions.

Electroencephalography (EEG) gauges a user's mental and emotional state. Emotional states like valence and arousal detected from EEG have been discussed as avenues for measuring usability \cite{gannouni2023}. To address this challenge, multiple integrated environments for interpretation and synchronization of multimodal data (e.g., EEG, audio, video, eye tracking, self-reported answers) were developed \cite{liapis2014, hussain2018}. Psychophysiological constructs such as valence can diverge from their psychometric (self-reported) counterparts to provide complementary information \cite{le-pailleur2020}. It has been demonstrated that universal emotions can be classified from EEG and facial expressions with reasonable accuracy \cite{matlovic2016}. Consequently, emotion detection can reliably indicate usability issues \cite{stefancova2018}. \citet{santos2022} further proposed a list of typical task and action-based usability smells that can be predicted from interaction logs and EEG.

Stress as a psychophysiological response has been predicted by using physiological signals, such as EEG and skin temperature \cite{liapis2021, liapis2021a}. Data from wearable sensors was used by \citet{maier2019} to detect states of stress, boredom and flow in games, the latter of which manifests when challenge corresponds to the user’s skill. 

\subsubsection{Automatic guideline evaluation}

This category, representing 13.5\% (n=21) of primary studies, focuses on validation of the quality of design by exploiting predetermined guidelines and heuristics as automated usability inspection. Typical approaches include source code analysis, interaction simulation and visual analysis of rendered user interfaces through image processing or computer vision.
Mirroring traditional usability inspection, techniques that automate heuristic usability evaluation of select design aspects (e.g., headings, graphical text, similarity between the homepage and other pages) lack the capacity to address factors that affect higher cognitive processes involved in human-computer interaction, such as decision-making and information needs \cite{dingli2014}. However, their value lies in providing immediate and early feedback to developers. 

Approaches based on source code have been explored, such as a reverse engineering framework used by \citet{almeida2015} to extract GUI behavior models for detection of usability smells. \citet{baek2016} proposed multi-level criteria for the generation of GUI models for testing, demonstrating that more fine-grained modeling is necessary for thorough coverage of flaws. To further account for dynamic aspects of user interfaces \citet{cassino2015} designed a tool that emulates human visual perception, models the hierarchical structure of the UI and simulates interactions.

Since automated UI testing tools such as Selenium or Puppeteer are commonly used by developers, \citet{marenkov2016, marenkov2016a} proposed a language based on Extensible Markup Language (XML) for the specification of usability guidelines for web application and a framework for their evaluation. Ontologies were later used for their expressive capabilities at capturing relations between concepts \cite{marenkov2017, robal2017}. For ontology-based evaluation of native desktop applications, \citet{meixner2021} created a plugin for an integrated development environment. Extensible analysis of the functional usability of mobile applications is possible with the framework presented by \citet{mathur2018} that decompiles the executable file to validate test cases. Evolutionary algorithms were used to also generate evaluation heuristics for UI aesthetic issues based on the context of user profiles \cite{soui2020}. Given the complexity of maintaining a high number of test cases, \citet{eskonen2020} proposed an automated solution that exploits deep reinforcement learning to efficiently explore GUIs.

Computer vision approaches were leveraged with deep learning for heuristic evaluation of screenshots and other images that capture user interfaces (e.g., thermostat screens \cite{ponce2018}). A few-shot learning framework for software UI testing was proposed by \citet{widodo2023} to classify 10 common UI flaws, while \citet{yang2021} sourced their guidelines from the rules of Material Design. Surface-level display flaws and discrepancies between implemented design and mock-ups were detected by \citet{liu2023} and \citet{moran2018}. Animations, which are traditionally challenging to analyze, were recorded and fed to unsupervised learning to detect anomalies that deviate from guidelines \cite{zhao2020}.

Natural Language Processing (NLP) can also contribute to guideline evaluation. \citet{hasanmansur2023} put forward a unified taxonomy of dark patterns in GUIs, then identified some of them automatically based on textual clues, as well as spatial and chromatic analysis. \citet{hsueh2024} demonstrated the potential of a Large Language Model for evaluating Nielsen’s heuristics based on user interaction scripts and screenshots.

\subsubsection{Usability issue encounter detection}

Research aimed at detecting precise instances when users encounter usability issues could be considered as the most pertinent to the subject of our literature review. It represents 11.6\% (n=18) of the primary studies. Typical approaches in this problem space involve interaction-based heuristics and usability smell detection, behavioral sequence matching, acoustic analysis, transcript generation and Natural Language Processing.

Rule-based approaches typically focus on heuristics and strategies for identifying pre-defined patterns that hinder user experience, which are sometimes systematically catalogued as usability smells. For web applications, tools were presented that log interaction events, identify smells (e.g., non-responsive element, freeform inputs for limited values) and offer suggestions for refactoring \cite{grigera2014, grigera2017, grigera2017a, ribeiro2019} while the toolkit by \citet{firmenich2019} supports automated A/B testing. \citet{vigo2017} concentrated on detecting adaptive behaviors in response to problematic website navigation, such as retracing and quick previews. Additionally, information architecture—a navigable information structure—can be validated through tree testing, a process that was standardized for remote automation by \citet{tapia2022}.

Due to the idiosyncrasies of mobile devices such as touch-based controls, a subset of rule-based usability issue detection is mobile-oriented. As a semi-automated approach for usability issue discovery, \citet{paterno2016} presented an interactive visualization for the exploration and comparison of timelines. \citet{goncalves2016} enhanced the automation by matching sequences of action events that are expected for executing tasks to sequences of actions performed by users, with mobile adaptations from a previous desktop-oriented solution. As an alternative to identifying usability issues based on the ground truth of correct task solutions, \citet{jeong2020} identified screens with usability issues based on collective inconsistency of user behavior. Usability smells that are notably pressing on mobile devices, such as links placed at high proximity and small text, were detected by \citet{paterno2017}.

Outside of traditional computing devices, \citet{benvenuti2021} presented a log-based behavioral sequence matching method based on Petri nets and trace alignment aimed at heuristically explaining usability issues of consumer electronics products. Closed sequential pattern mining was used by \citet{jorritsma2016} for identifying usability issues based on frequent behavioral patterns in radiology systems. Similarly in VR, \citet{harms2019} detected important tasks by analyzing task trees from actual use data, along with pertinent usability smells. VR headsets also provide embedded signal sources such as EEG, as well as head and hand gesture tracking, which \citet{kaminska2022} leveraged in machine learning to predict the presence of usability issues.

Due to the generic nature of rule-based detection, its primary obstacle rests in the inability to capture more complex and nuanced issues that require human reasoning \cite{grigera2017a}. Therefore, to reveal such patterns, more recent approaches pursued the analysis of speech (think-aloud) and video. \citet{fan2019} examined the link between usability issues and features of verbalizations, including acoustics (e.g., pitch) and transcript coding. By leveraging these features, prediction of usability issues can achieve better performance \cite{fan2020}, making it useful for UX professionals when incorporated into a visual analytics tool \cite{fan2020a}. Continuation of this research notes a change in paradigm to AI-assisted discovery of usability issues (see \ref{sec:research-assistants} \hyperref[sec:research-assistants]{Research assistants}).

\subsubsection{Feedback evaluation}

Despite explicit feedback potentially exhibiting a number of potential biases (e.g., recall, social desirability), it is nevertheless a valuable source of information about usability issues. Therefore, 7.7\% (n=12) of primary studies are devoted to processing of explicit feedback data to extract attributes such as topics, sentiment and salience, often in coordination.

Topic modeling techniques like Latent Dirichlet Allocation (LDA) can reveal underlying themes in feedback, supporting its summarization \cite{guzman2014, sadigov2024} and uncovering user-reported usability issues \cite{eiband2019, eiband2020}. Classification algorithms can be effective for pre-defined categorization of feedback, such as into groups oriented at specific stakeholders \cite{maalej2016}, according to a usability issue taxonomy \cite{khalajzadeh2023} or anatomical structures of arguments \cite{wang2020}. \citet{bakiu2017} extracted feedback about specific software features by applying collocation algorithms, an approach potentially challenged by inconsistent wording. For emergent topic generation, \citet{asnawi2023} proposed an innovative topic modeling technique that accounts for inter-topic dependencies with a model that reflects dependencies between tokens, achieving higher coherence than previous approaches.

Sentiment classification performed using traditional Natural Language Processing (NLP) techniques like lexical analysis was used to assign polarity to the observations and attitudes expressed in feedback, such as reviews \cite{guzman2014}. Instruments for sentiment analysis were also used to enhance topic modeling techniques \cite{eiband2019, eiband2020}. More recently, novel deep-learning-based approaches alongside commercial NLP solutions were assessed by \citet{sanchis-font2019, sanchis-font2021}, demonstrating the potential for classifying positive and negative sentiment, as well as the challenge of accurately detecting neutrality.

Since providing valuable feedback can be challenging, \citet{krause2017} proposed a method for evaluating the helpfulness of user feedback. An intervention where relevant style guidelines are displayed to users allows them to revise their feedback, thereby improving its quality.

\subsubsection{Research assistants}
\label{sec:research-assistants}

A budding field of study at 7.1\% (n=11) is concerned with the creation of conversational research assistants and other assistive technologies. These assistants can aid researchers with usability evaluation, or enhance usability research by adaptive interaction with participants.

In an earlier work that predated broader adoption of Large Language Models (LLMs), which ignited greater interest in research assistants, \citet{kim2019} investigated whether the framing of a survey within a chatbot interaction could mitigate satisficing—low-effort responding in online usability surveys. In a flow-based interaction, they found that chatbots, especially when communicating in a familiar tone, can lower satisficing through stronger engagement. Participants also show some preference for conversational surveys \cite{celino2020}.

Besides improving engagement, conversational assistants are investigated as a potential solution for emulating the strengths of moderated UX research within unmoderated research techniques, by imitating some capabilities of human moderators. \citet{liu2024} designed a Hybrid UI (a combination of Graphical and rule-based Conversational UI) for automation of structured  Repertory Grid Technique interviews, demonstrating potential despite ecological validity limitations due to an artificial setting. \citet{kuric2024} examined the ability of GPT-4 to ask follow-up questions during a usability test. Based on the identified flaws, they classified the types of context that a system for generating reasonable follow-up questions should incorporate. 

Originally, the assistance of AI in research data analysis was viewed as the application of machine learning to predict relevant patterns in behavior, speech, posture, etc. These features were then visually highlighted as potential indicators of problems for UX professionals \cite{soure2022, batch2024}. While this description is still accurate, advances in Natural Language Processing and Conversational User Interfaces have led to a shift to more interactive conversational AI assistants. For example, \citet{bisante2024} presented a tool that employs GPT-4 to guide novice designers in the process of cognitive walkthrough. Exploratory evaluation showed that its suggestions aligned with expert evaluations, although the tool ignored a number of visual issues that humans found evident. While participants rated the solution positively, they also encountered errors, hallucinations and trust issues.

Multiple Wizard-of-Oz studies simulated an AI assistant during the process of analyzing usability testing recordings \cite{fan2022, kuang2023a, kuang2024}. Text and voice modalities were found to be beneficial for distinct reasons, since researchers used them to ask different types of questions \cite{kuang2023a}. Aside from questions about user actions or mental models, the questions implied interest in functionality such as design suggestions, note taking, voice control, search of the Web and access to contextual information (e.g., demographics of the participant, task details). In a later follow-up, \citet{kuang2024} also used an LLM (ChatGPT) to generate suggestions from transcripts, then explored the timing and manner in which researchers interacted with them. Their results align with the direction of AI as an augmentative tool that is subject to human expertise, demonstrating limitations in AI-identified usability issues and a preference for researchers to prioritize their own analytical skill. Explanations play a critical role for building trust towards suggestions, but they also enhance the risk of superficial trust, given that some explanations may collapse under deeper scrutiny \cite{fan2022}.

\subsubsection{Visual complexity evaluation}

Visual information in user interfaces can be challenging to cognitively process. Higher complexity results in usability issues. To support monitoring and reduction of users’ cognitive load, automated analysis of visual complexity was at the center of 3.2\% (n=5) of studies. Visual complexity analysis typically involves images that faithfully represent the UI seen from users’ perspective.

As exemplified in the evaluation tool by \citet{oulasvirta2018}, automated methods for evaluation of visual complexity represent a continuation of previous approaches for quantifying visual complexity through measures such as amount of information, visual clutter, contrast or symmetry. To process input images, \citet{bakaev2018, bakaev2018a} extracted rectangular areas, identified text with deep-learning optical character recognition (OCR) and visual elements by using classification with histogram-based feature extraction. For calculated metrics such as the number of all UI elements or the area under text elements, they demonstrated correlations with user-reported counterparts. They proposed the index of visual complexity, a derived metric with significant correlation with perceived complexity. Concentrating on metrics that are more straightforward to measure and calculate, \citet{boychuk2019} assessed the correlation between visual complexity and measures of JPEG and PNG file size and information entropy, demonstrating an improved predictive accuracy in multifactor regression.

\citet{miniukovich2018} explored the effects of the type of stimulus (website, book page) and dyslexia, on which visual factors affect perceived visual complexity. Dyslexics did not differ from typical users in factors that increase their cognitive load. Differences between book pages and websites implied a difference in user expectations.

\subsubsection{Automated feedback generation}

Modeling and synthesis of human-like feedback, such as preferences, responses to questions and qualitative assessments, is a nascent field of study covered by 3.2\% (n=5) of the corpus. Feedback generated by AI as a simulation of participants in usability testing is a controversial topic, raising concerns about the AI model’s invisible biases, ability to reflect user characteristics, and emulate realistic perception and thinking \cite{xiang2024}.

Subjective preferences about the appearance of UI design elements were the subject of prediction by \citet{zhou2020}. The presented framework collects user feedback, then leverages collective learning to predict a design with optimal user preferences.

Large Language Models (LLMs) were utilized to generate textual feedback at different design stages. For mockups, \citet{duan2023} introduced a plugin for the Figma design tool, which generates constructive suggestions as a reflection of compliance to design guidelines. The characteristics and the contributory value of obtained feedback was not yet assessed. \citet{xiang2024} proposed a tool leveraging Chain-of-Thought (CoT), where two agents (representing the app and the user respectively), establish user expectations, simulate perception, and their interaction with the user interface. Usability issues were coded in five heuristic categories based on the misalignment between expectations and the user interface. Despite yielding some potentially useful information, results revealed incomplete feedback, with discrepancies from feedback provided by actual users. The LLM feedback also lacked the ability to reflect user characteristics defined by their needs and experiences. In the context of surveys and games, similar conclusions were reached by \citet{hämäläinen2022, hämäläinen2023}, who also identified the risk for the abuse of LLMs for generation of fake AI responses in crowdsourcing.

\subsubsection{Aesthetics evaluation}

Aesthetic characteristics are often considered hedonic, yet their contribution to the attractiveness and perception of user interfaces makes them integral to usability. A small 2.6\% segment of studies (n=4) focused on the evaluation of aesthetics with computer vision, image processing and eye tracking.

Convolutional Neural Networks (CNNs) were used by \citet{dou2019} to regress the aesthetic of website UIs with high correlation to aesthetic ratings by actual users. While a step forward from models based on manually-created spatial and chromatic metrics, some inconsistencies with real user ratings were still present. These can be attributed to top-down factors (e.g., user expectations) or high-level design concepts that are not captured by traditional metrics. \citet{xing2021} also exploited a larger dataset of GUI images from social media, adopting engagement metrics such as likes as predictors of aesthetics. However, this research does not address exposure and popularity bias, nor the potential effects of other variables present in the context of a social media post.

In a gaze-tracking-driven approach that takes into account the visual attention of users in the actual environment of the evaluated UI, \citet{gu2021} differentiated between web pages with good and bad aesthetics based on the introduced index of visual attention entropy. Their findings lend credibility to the hypothesis that aesthetically pleasing experiences are perceived more fluently.

\subsubsection{Perceived affordance evaluation}

In design, affordance determines the means by which a system can be used \cite{amri2020}. In the context of usability issue detection, 2.6\% (n=4) of relevant works examined properties of GUI elements that influence how users perceive interactive potential, such as saliency or clickability/tappability. Visual attention was also studied, with saliency representing only its bottom-up component, while top-down factors further shape perception.

In an earlier study, \citet{koch2016} applied Gestalt Laws of perception (e.g., proximity, similarity) to heuristically identify visual associations between GUI elements. In a more data-driven approach, \citet{xu2016} used machine learning to predict visual attention based on the features of user interface elements and mouse and keyboard interactions. Attention maps generated in simple self-contained text-editing tasks were more similar to the eye tracking ground truth than previous saliency map solutions, while also enabling dynamic attention prediction.

Deep learning has enabled more complex analysis of patterns in GUI images. Recall of gaze was predicted as a proxy of saliency in web page screenshots \cite{cheng2023}. To automatically predict the perceived tappability of elements in mobile GUI interfaces, \citet{schoop2022} proposed a neural network model and adopted techniques of Explainable Artificial Intelligence (XAI) to justify its predictions. Region-based explanations on their own were found as not granular and specific enough to enable meaningful interpretation of the root cause of mismatched perception.

\subsection{Technology (RQ3)}
\label{sec:technology}

\begin{figure}[!ht]
  \centering
  \includegraphics[width=0.6\linewidth]{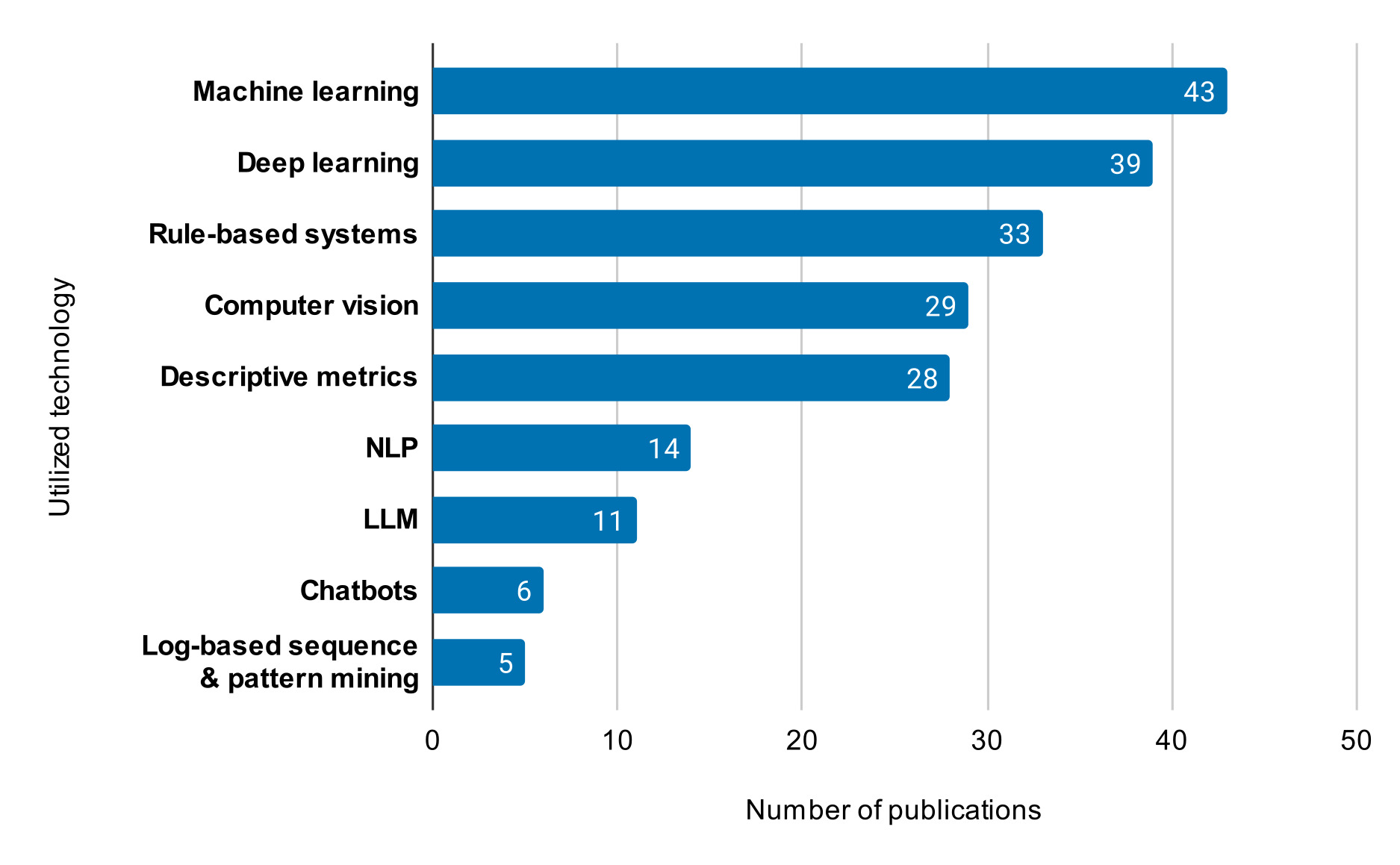}
  \caption{Technological paradigms in automated usability issue detection. The most common technologies are Machine learning and Deep learning, followed by Rule-based systems.}
  \Description{The image is a horizontal bar chart displaying the number of publications utilizing specific technology paradigms. The horizontal axis represents the number of publications while the vertical axis represents the technology paradigm utilized. The most common paradigm is Machine learning with 43 publications, then deep learning with 39, rule-based systems with 33, computer vision with 29, descriptive metrics with 28, followed by NLP, LLM, ChatBots and Log-based sequence & pattern mining.}
  \label{fig:tech-bar}
\end{figure}

\begin{figure}[!ht]
  \centering
  \includegraphics[width=0.6\linewidth]{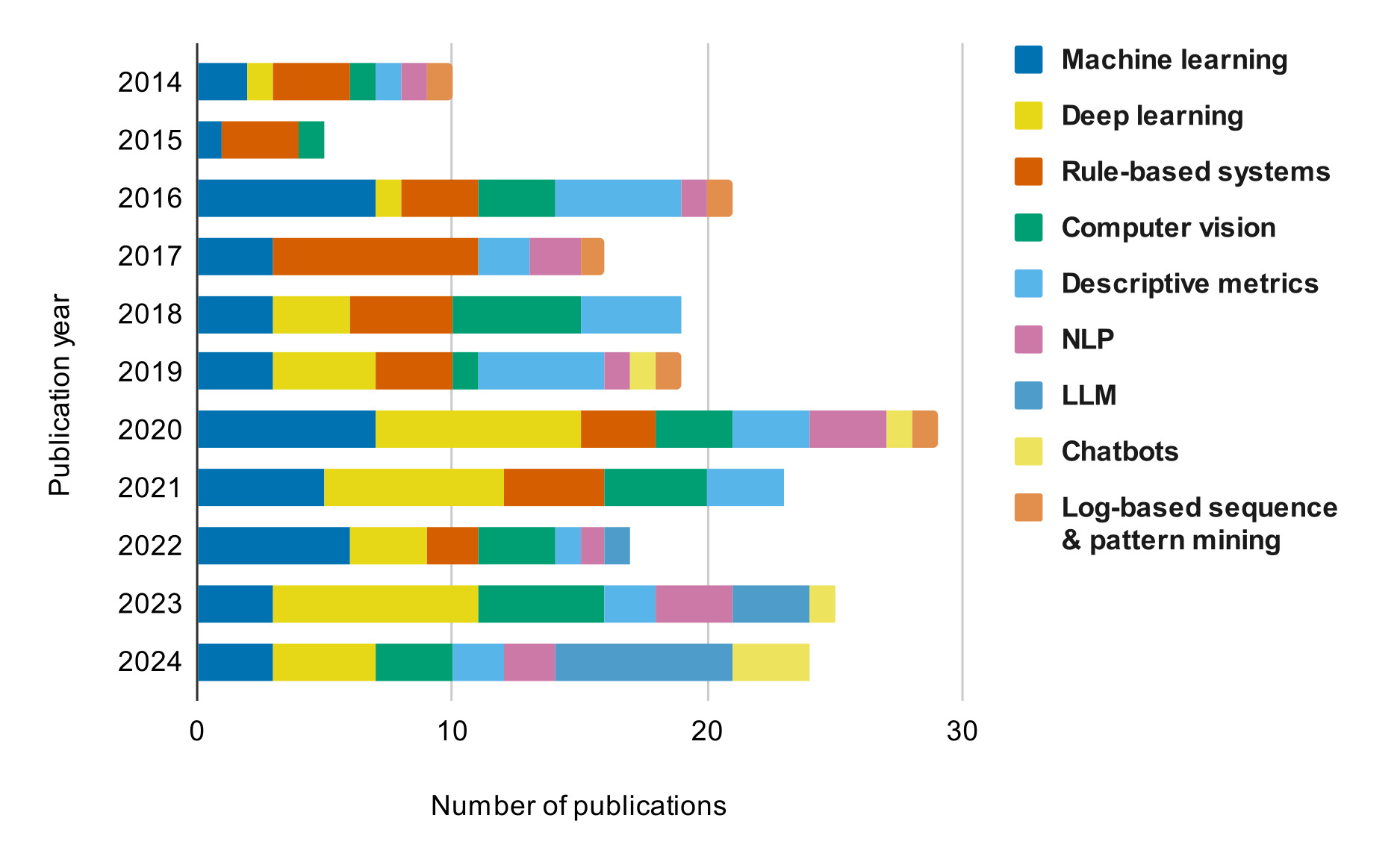}
  \caption{Per-year distribution of technological paradigms in usability issue detection. While paradigms have remained relatively stable over time, the popularity of LLMs and Chatbots has increased in recent years.}
  \Description{The image is a horizontal stacked bar chart displaying the proportions of technology paradigms across publications for each year. The horizontal axis represents the number of publications while the vertical axis represents the publication year from 2014 to 2024. Proportions of technological paradigms, displayed as different colored segments in each bar demonstrate consistent presence of machine learning, while rule-based systems are on the decline. In the most recent years, LLMs and Chatbots have significantly grown in prevalence.}
  \label{fig:tech-years}
\end{figure}

The utilization of technologies that form the foundation of automated detection of usability issues is summarized in \autoref{fig:tech-bar}. Annual distribution is expanded upon in \autoref{fig:tech-years}. Considering the hierarchical relationships in technology (e.g., deep learning as a subset of machine learning, LLM as an NLP model), to maintain labels with distinct meanings, supercategories were only assigned to methods and techniques distinct from their subcategories. For example, a study from the 'deep learning' category was only given the second ‘machine learning’ label if it also included traditional machine learning techniques.

The undisputed expressiveness and adaptability of machine and deep learning for analysis of complex patterns have resulted in their application to a wide palette of discriminative and generative problems \cite{souza2022, wang2020, issa2020, xing2021, cheng2023, yang2020, yang2021}. In recent years, deep learning has been outweighing traditional machine learning approaches. However, traditional techniques have not been substituted completely, owing to the advantages they offer in speed, flexibility and interpretability \cite{santos2022}. These aspects can be crucial in research aimed at justifiably diagnosing usability issue findings. Being reliant on training data that reflects latent patterns, machine and deep learning can—despite its advantages—still be subject to validity threats. For example, when predicting saliency based on recalled gaze position \cite{cheng2023}, an angular error of 2-5° between the recalled and actual gaze position is substantial in the context of visual angles \citep{kuric2025}, raising some concerns about bias in the recall data and the predicted constructs.

Systems based on hand-crafted rules are common, albeit declining in popularity, becoming completely absent during the last two years of the surveyed period. Typically, they are used for evaluation of usability measures \cite{verkijika2018, yu2018, asemi2022} or heuristics \cite{mathur2018, robal2017, soui2020} based on source code, interaction logs or simulations \cite{cassino2015, marenkov2017}.

Computer vision and adjacent image processing techniques are frequently used in tandem with machine and deep learning, either through hand-crafted visual features or in CNNs. Characteristic uses include support of automated usability testing by capturing the subject’s affective state \cite{feijofilho2016, desolda2021, giroux2021} and analysis of the appearance of user interfaces, either to heuristically identify potential issues \cite{batch2024, hsueh2024, widodo2023, liu2023} or analyze visual properties \cite{bakaev2018a, schoop2022, dou2019}.

Research centered around the calculation of descriptive metrics typically introduces constructs aimed at reflecting usability issues, methods and tools for their automated measurement \cite{blanco-gonzalo2014, salomon2023, tapia2022, fan2019, oulasvirta2018}, or explores their correlation to usability-related constructs \cite{katerina2018, harrati2016}.

Language, spoken or written, is the most explicit and expressive means for users to communicate their internal experience. Therefore, Natural Language Processing (NLP) fills a key niche in automating the evaluation of feedback obtained from usability testing \cite{krause2017, fan2020, fan2020a, soure2022, batch2024}, reviews \cite{bakiu2017, guzman2014, maalej2016, eiband2020, eiband2019} and social media posts \cite{sadigov2024}. Large Language Models (LLMs), corresponding to their conversational aptitude and ability to identify relationships between concepts, have been used primarily for automated generation of usability findings \cite{duan2023, xiang2024, hämäläinen2022, hämäläinen2023} or to design conversational assistants \cite{kuang2024, liu2024, bisante2024, kuric2024}, but also to extract topics from feedback \cite{asnawi2023} and perform heuristic evaluation \cite{hsueh2024}.

Chatbots leveraged in research assistants are either rule-based \cite{kim2019, celino2020} or LLM-based \cite{kuang2024, liu2024, bisante2024}. Not all chatbot studies implemented a specific technology, as is the case for \citet{kuang2023a}, a Wizard-of-Oz study that investigated user expectations and interactions with simulated AI chatbots.
Sequence and pattern mining techniques were utilized primarily to identify usability issue encounters from interaction data \cite{harms2019, jeong2020, paterno2017, jorritsma2016}.

\subsection{Artificial intelligence (RQ4)}

\begin{figure}[!ht]
  \centering
  \includegraphics[width=0.6\linewidth]{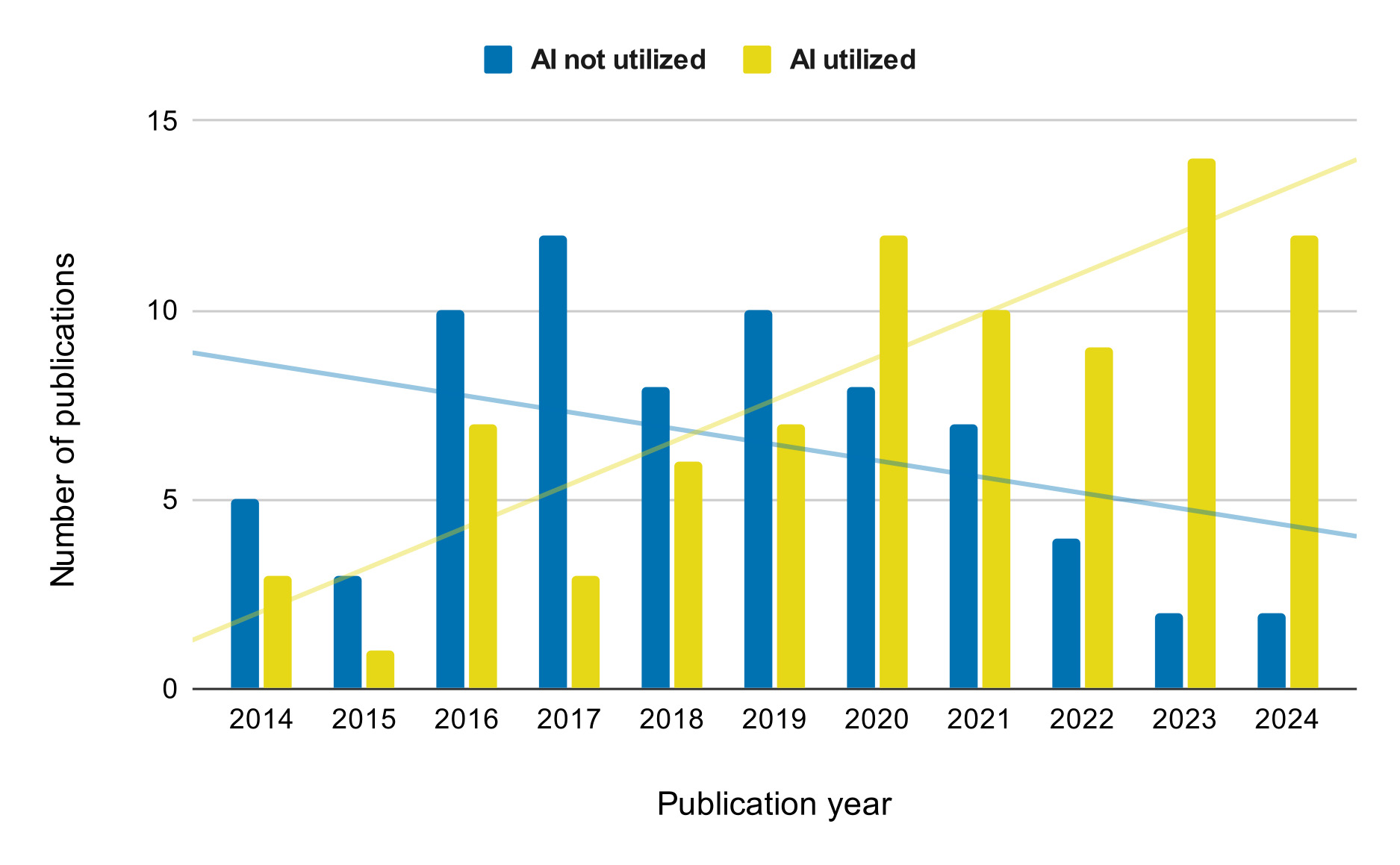}
  \caption{Per-year distribution of studies based on their implementation of AI. There is a notable increase in AI-focused publications across years while publications without any form of AI are steadily declining.}
  \Description{The image is a vertical bar chart displaying the number of publications with and without AI across years (two bars for each year). The horizontal axis represents the publication year from 2014 to 2024 while the vertical axis represents the number of publications. The number of publications utilizing AI grows over the years, with a minimum of 1 publication in 2015 and reaching a peak of 14 publications in 2023. On the other hand, publications that do not utilize AI decrease, with the maximum of 12 in 2017 and minimum of 2 in 2024.}
  \label{fig:ai-years}
\end{figure}

For its focus to match the definition of AI, a study needs to include at least one of the three following technology labels established in \ref{sec:technology} \hyperref[sec:technology]{Technology}: Machine Learning, Deep Learning or LLM. Over the last decade, there is a balance between studies that focus on some form of AI (84 publications, 54\%) and those that do not (71 publications, 46\%), although interest is already gradually shifting in favor of AI. As pictured in \autoref{fig:ai-years}, non-AI studies aimed at automation of usability issue detection are on the decline and studies involving AI are on the rise. There is a strong correlation between the year and number of publications utilizing AI, suggesting significant growth ($r_s(10)=.88, p<.001$). LLMs in particular can be reasonably expected to further trend as the locus of attention, given that articles involving them have 17 citations per year on average on Google Scholar (for comparison, both machine learning and deep learning have 11 on average).

\subsection{Devices (RQ5)}

\begin{figure}[!ht]
  \centering
  \includegraphics[width=0.6\linewidth]{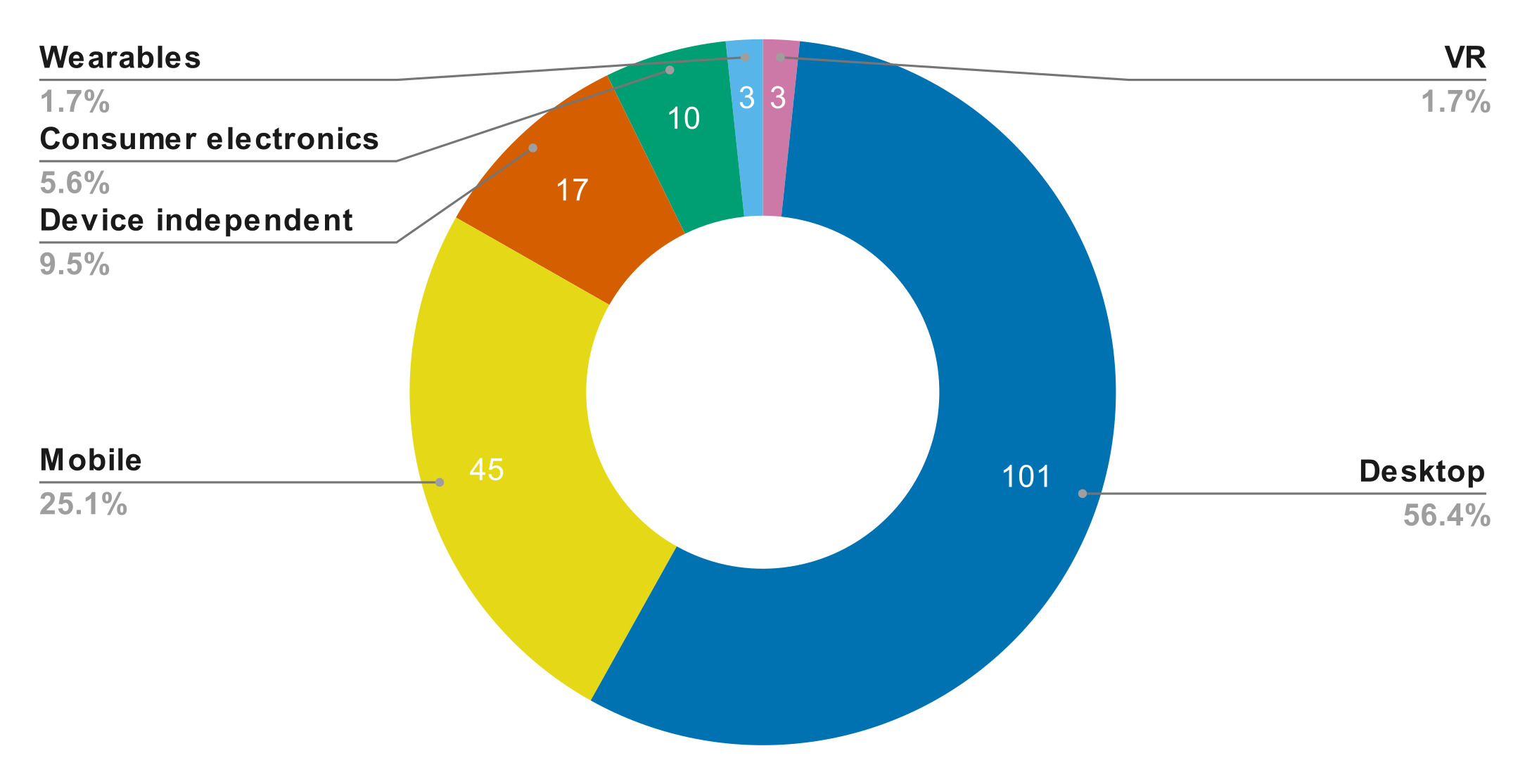}
  \caption{Distribution of the types of devices for which usability issue detection methods were validated. The most common devices are Desktop and Mobile.}
  \Description{The image is a pie chart displaying the proportion of types of devices used for  validating the issue detection method. The largest segment involves Desktop devices with 101 publications (or 56\%), Mobile devices with 45 publications (or 25\%) and Device independent approaches with 17 publications (or 10\%). Additional segments in order include VR, Wearables and Consumer electronics.}
  \label{fig:dev-pie}
\end{figure}

\begin{figure}[!ht]
  \centering
  \includegraphics[width=0.6\linewidth]{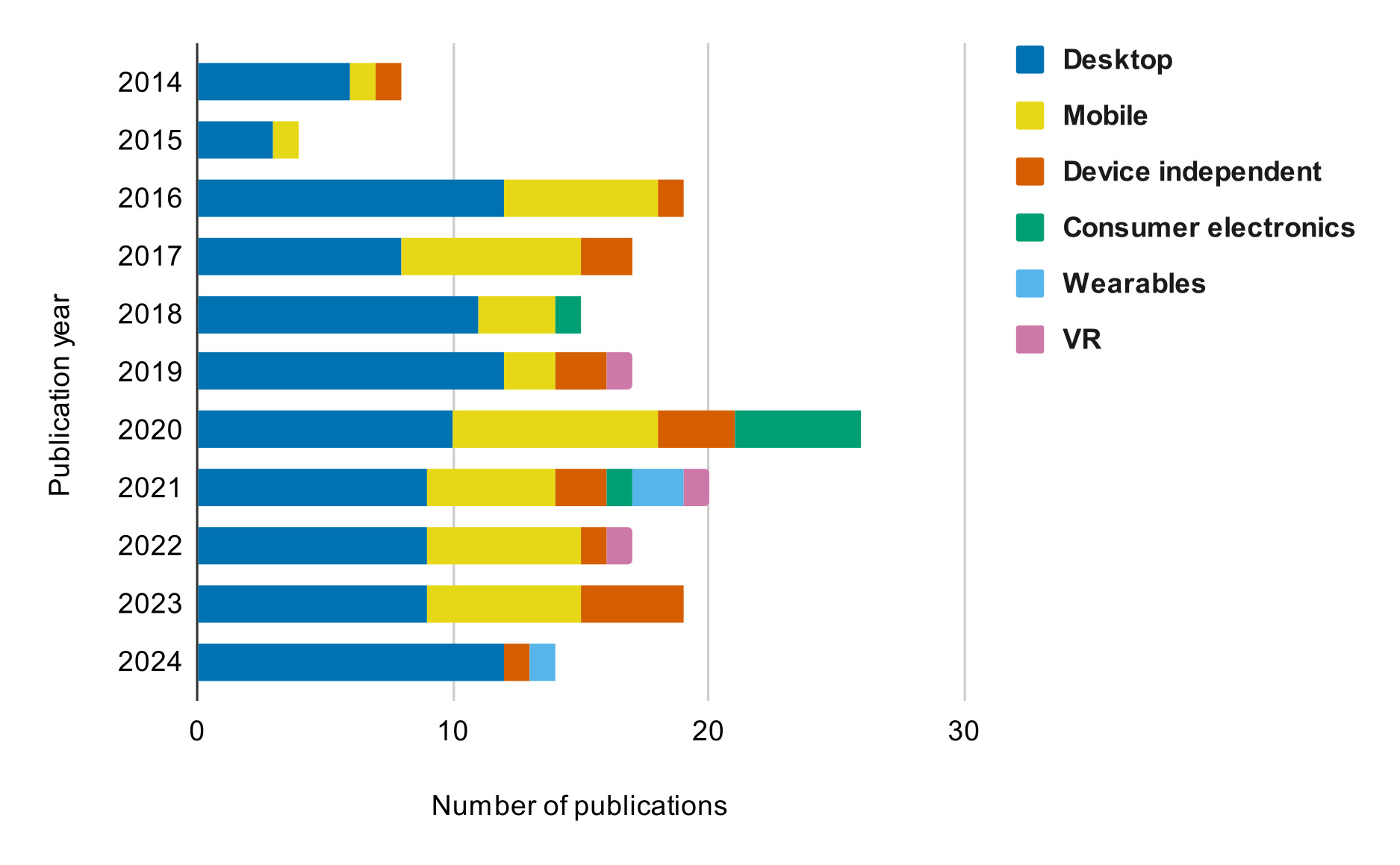}
  \caption{Per-year distribution of the types of devices for which usability issue detection methods were validated. The distributions of devices across years are mostly consistent, although there were no mobile-focused studies in 2024, in spite of Mobile being a prevalent category in previous years.}
  \Description{The image is a horizontal stacked bar chart displaying the distributions of devices across publications for each year. The horizontal axis represents the number of publications while the vertical axis represents the publication year from 2014 to 2024. Distributions of devices are displayed as different colored segments in each bar. Two most prevalent devices, Desktop and Mobile, are consistent across years.}
  \label{fig:dev-years}
\end{figure}

An overview of the types of devices targeted by automated usability issue detection is provided in \autoref{fig:dev-pie}. Desktop devices are the most common and can be effectively viewed as the default, given that researchers seldom identify desktop computers as a key focus like they do with mobile devices \cite{mathur2018, moran2018, soui2020, dahri2019}. Besides mobile devices becoming more prevalent in daily use, they can also be attributed a number of specific challenges, such as display issues due to hardware and configuration variety \cite{liu2023}, touch and gesture controls \cite{schoop2022, kiseleva2016}, and dynamic environments that can create distracting or stressful conditions \cite{feijofilho2016, blanco-gonzalo2014}. The distribution between studies targeting mobile and desktop devices is also rationally stable over time (see \autoref{fig:dev-years}).

Device-independent techniques are typically integrated in a context where the devices are of limited significance, such as analysis of speech \cite{chen2021} or textual feedback \cite{sanchis-font2019, sanchis-font2021, khalajzadeh2023, wang2020}.

Consumer electronics such as televisions, remote controls, microwaves and other everyday items have so far been involved in automated usability evaluation only rarely \cite{ponce2018, bures2020, benvenuti2021, fan2020, fan2020a}. Nevertheless, it can be challenging to make use of their full functionality without reading a manual, making their complex user interfaces a domain in need of further study. 

Wearables and VR headsets are emergent devices that present new ways for humans to interact with computers. So far, the number of studies in the context of our investigated topic and VR has been limited \cite{harms2019}. The research aimed at automated usability evaluation with VR and wearables was primarily focused on investigating the potential of built-in physiological sensors to infer the subject’s affective state and user experience \cite{liapis2021a, holzwarth2021, kaminska2022}. Significantly, smartwatch apps were also the subject of a study that investigated LLM-driven simulation of usability feedback \cite{xiang2024}.

\subsection{Data (RQ6)}

\begin{figure}[!ht]
  \centering
  \includegraphics[width=0.6\linewidth]{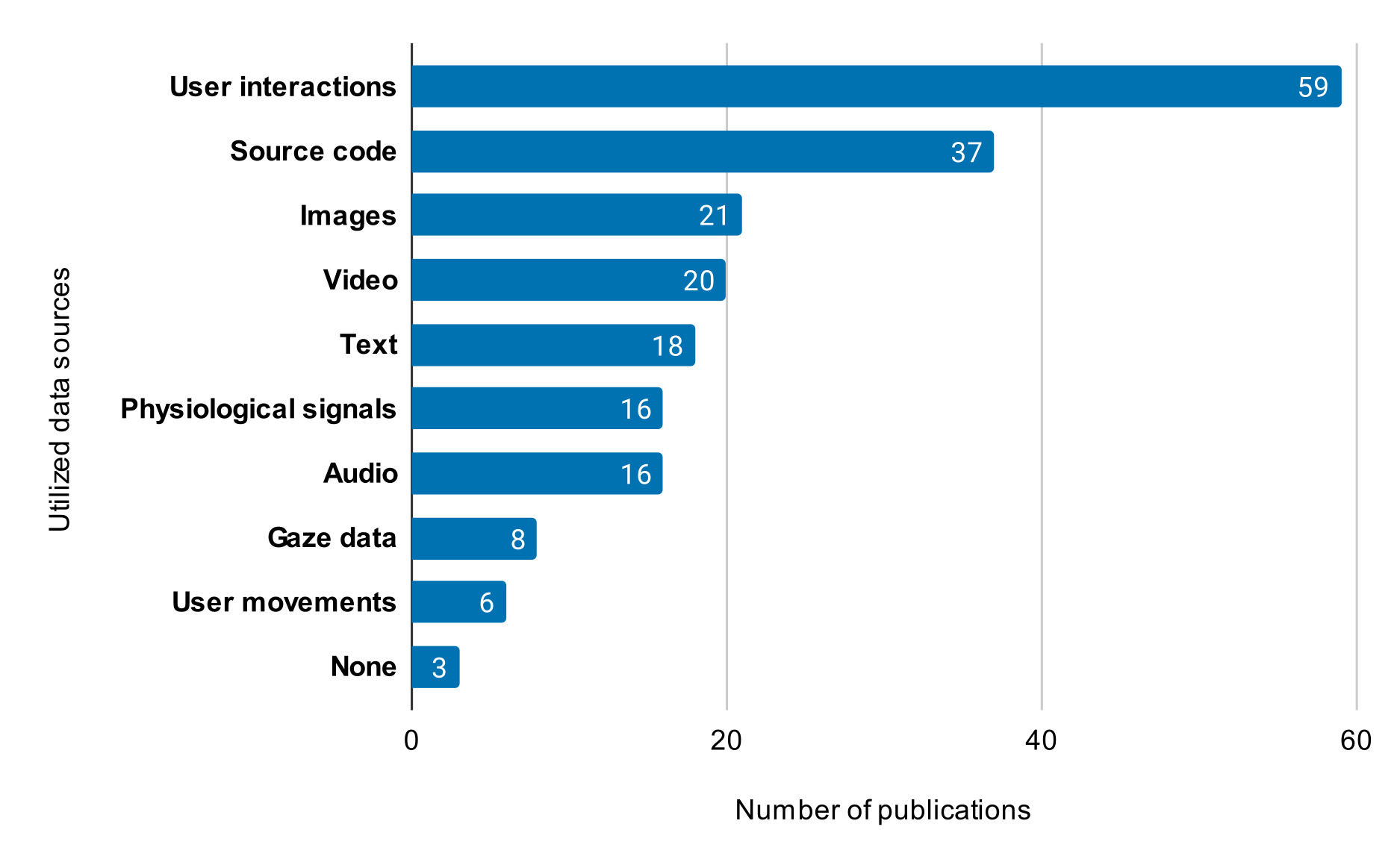}
  \caption{Types of data used in automated usability issue detection. The most common data types are User interactions and Source code, followed by Images, Video, and Text.}
  \Description{The image is a horizontal bar chart displaying the number of publications utilizing specific data sources. The horizontal axis represents the number of publications while the vertical axis represents the utilized data source. The most common data sources include User interactions with 59 publications, then Source code with 37, Images with 21, Video with 20, Text with 18, followed by Physiological signals, Audio, Gaze data and User movements. In three publications, no data was utilized for actual identification automation purposes.}
  \label{fig:data}
\end{figure}

Automated usability issue detection is contingent on data that retains essential indicators by which room for improving usability can be identified. The types of data from which literature infers usability problems are varied (see \autoref{fig:data}). Some studies explore a singular data source, although multimodal approaches offer a more comprehensive view by integrating diverse data sources. Multimodal approaches not only explore relationships between variables to improve the understanding of their interactions, but also enhance predictive capacity \cite{salomon2023, poore2017, hussain2018, santos2022, kiseleva2016, le-pailleur2020, razzaq2023}.

User interactions can comprise logs of high-granularity user actions (e.g., move movements, keystrokes, clicks/taps, physiological responses) \cite{batch2024}, their less granular semantic interpretations (e.g., dwell time on GUI elements, mouse movement patterns) \cite{gardey2022, katerina2018} and task-oriented observations (e.g., completion time, task traversal) \cite{villamane2024}. The position of interactions as the most prevalent source of data can be attributed to their status as implicit behavioral signals. This makes them non-intrusive and easy to collect, whether during usability testing or normal usage \cite{jorritsma2016, vigo2017}.

Source code data is used in approaches that rely on heuristic analysis of code, structure of GUI and its components, in some instances simulating user interactions \cite{cassino2015, marenkov2016} or generating usability issues with AI \cite{duan2023, xiang2024}. A common framing for using source code is for quick, early usability feedback for developers \cite{marenkov2017, grigera2017}, or to provide easy-to-use summative techniques for evaluation of existing systems \cite{verkijika2018}. Due to the costs of iterative usability testing, some authors have argued in favor of source-code-based usability inspection during development \cite{robal2017}. However, development of source code is itself costly. Therefore, whenever possible, most salient usability issues should be resolved before they need to be fixed in code. In this context, source code inspection can help mitigate usability issues and aesthetic flaws introduced by developers as they are implementing designs from GUI mockups and prototypes \cite{moran2018, soui2020}.

In all studies that leveraged image data, its contents captured the appearance of GUIs or visual components as they are realistically perceived by users \cite{hasanmansur2023, hsueh2024, widodo2023, boychuk2019, schoop2022}. By contrast, screen capture video—commonly captured to be manually analyzed alongside semi-automated usability evaluation results \cite{fan2019, batch2024, kuang2023a}—was rarely utilized as a data source for automated techniques. Besides its use for validating animations \cite{zhao2020}, this can be attributed to the complexity of video analysis. Instead, user-facing cameras were the main source of video data, enabling the possibility of analyzing facial expression and body language as reflections of affective and cognitive states \cite{desolda2021, stefancova2018, matlovic2016, georges2016}. 

Text as a source of data appears almost exclusively in studies dedicated to extracting significant information from explicit feedback \cite{bakiu2017, khalajzadeh2023, sanchis-font2021, sanchis-font2019, eiband2020, eiband2019}, typically with techniques of machine learning and Natural Language Processing.

Physiological signals (e.g., electroencephalography, electrodermal activity, electrocardiography) were used—sometimes alongside facial expressions or user interactions—to infer affective state \cite{stefancova2018, matlovic2016, liapis2014, santos2022, le-pailleur2020}, or to evaluate user experience factors such as engagement and cognitive workload \cite{yu2018, frey2016, aleksander2018}.

Audio signals have been processed as indicators of the relationship between emotions and user experience \cite{issa2020, soleimani2017} or to otherwise analyze speech cues of encountered usability issues in think-aloud sessions \cite{fan2020, fan2020a, fan2019}. Voice modality was also found as key for the development of conversational research assistants \cite{kuang2023a}.

Sensors for inferring body movements, such as 3D body meshes from Kinect or head pose from VR devices, have been used to evaluate usability in immersive environments (VR, games) and to map users’ emotional states \cite{holzwarth2021, kang2014, razzaq2020}. Gaze tracking capable of tracing a person’s visual attention on the screen is a standard technique for exploring patterns by which users mentally process digital environments \cite{liapis2014, hussain2018}. In automated usability assessment, it has been used to link emotional states to on-screen objects \cite{georges2016}, calculate factors that affect cognitive processing such as task load and visual attention entropy \cite{shojaeizadeh2019, gu2021} and as part of multimodal user experience and engagement evaluation approaches \cite{poore2017, souza2022}.

A small number of conceptual studies did not operate with data that was intrinsically descriptive of systems, GUIs or user attitudes and behaviors towards them in the typical sense. For illustration, \citet{hämäläinen2022, hämäläinen2023} studied the ability of GPT-3 to generate humanlike survey answers, based only on a prompt and the language model’s vast dataset of text from the Internet. A conceptual framework was presented by \citet{gupta2023}, where AI personas assess the usability of an ontological representation of software requirements specifications.

\subsection{Maturity (RQ7)}

\begin{figure}[!ht]
  \centering
  \includegraphics[width=0.6\linewidth]{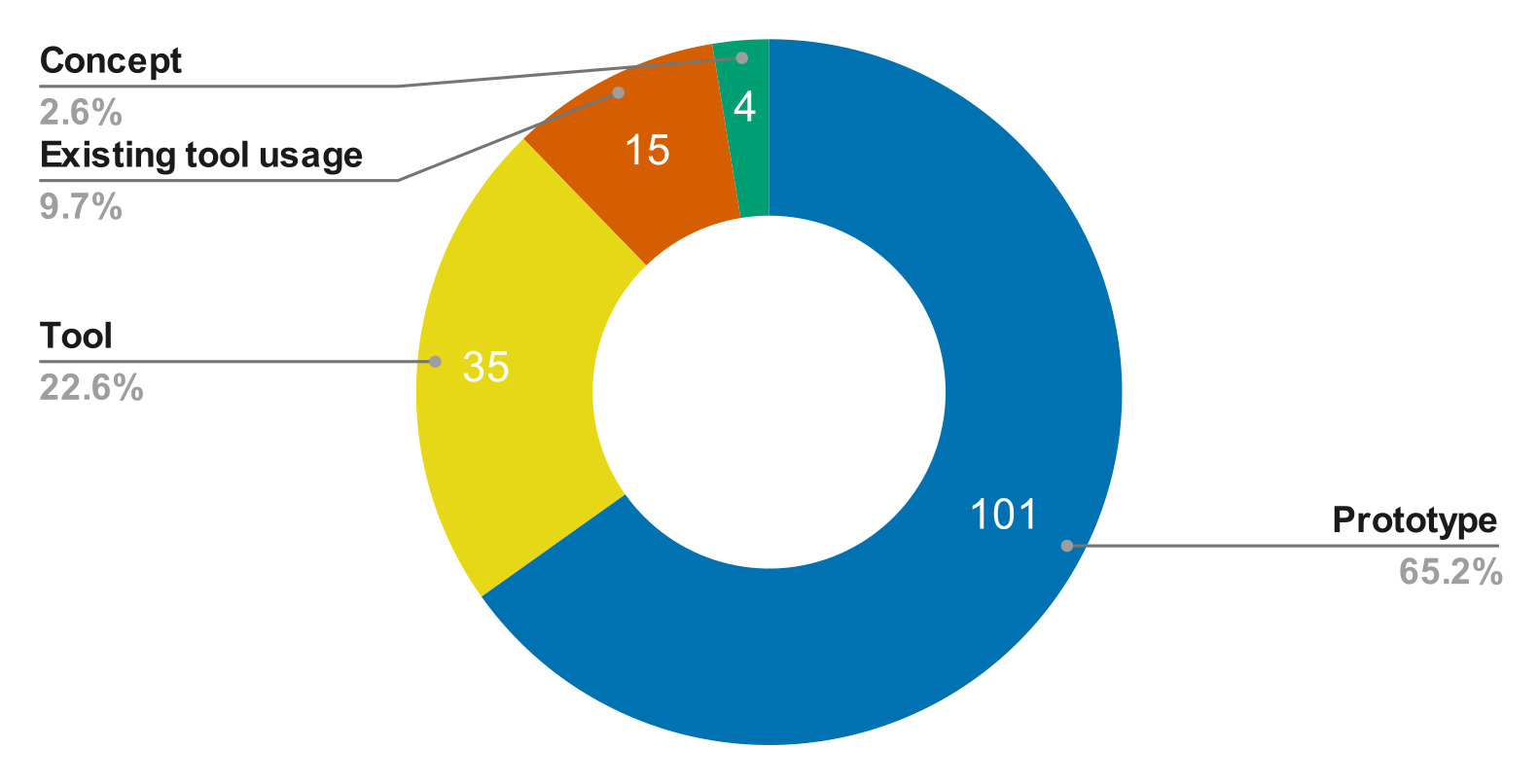}
  \caption{Maturity of automated usability issue detection approaches in research. The majority (65\%) of approaches are in the prototype stage, with fewer providing completed tools (23\%).}
  \Description{The image is a pie chart displaying the maturity of automated usability issue detection methods. The largest segments of studies involve Prototypes with 101 publications (65\%) and Tools with 35 publications (23\%). Additional segments include Existing tool usage with 15 publications and Concepts with 4 publications.}
  \label{fig:maturity}
\end{figure}

Primary research of automated detection of usability issues can be divided into four categories depending on the degree of its maturity (see \autoref{fig:maturity}). Research at the concept stage typically proposes a framework or a method without its actual implementation and validation. Only a small number (n=4) of publications falls under this category, two of them encompassing simulated (AI) user feedback \cite{hämäläinen2022, gupta2023} while the others cover automation in tree testing \cite{tapia2022} and A/B testing \cite{cruzgardey2020}.

The majority of primary research (n=101) is in the prototype stage, proposing a framework, technique or a model that the authors proceed to empirically evaluate by performing experiments and case studies. The output of a smaller number of publications (n=35) takes this a step further by presenting a ready-to-use tool. Finally, studies that investigate existing tools comprise their own category (n=15). While \citet{namoun2021} compared (and criticized) web usability evaluation tools, other studies focused primarily on benchmarking and ranking systems, assessing compliance and discovery of usability issues in specific domains (e.g., e-government websites in in a specific region, such as the sub-Saharan Africa) \cite{verkijika2018, csontos2021, ismailova2017, zarish2019, parajuli2020}.

\subsection{Participant involvement}

\begin{figure}[!ht]
  \centering
  \includegraphics[width=0.6\linewidth]{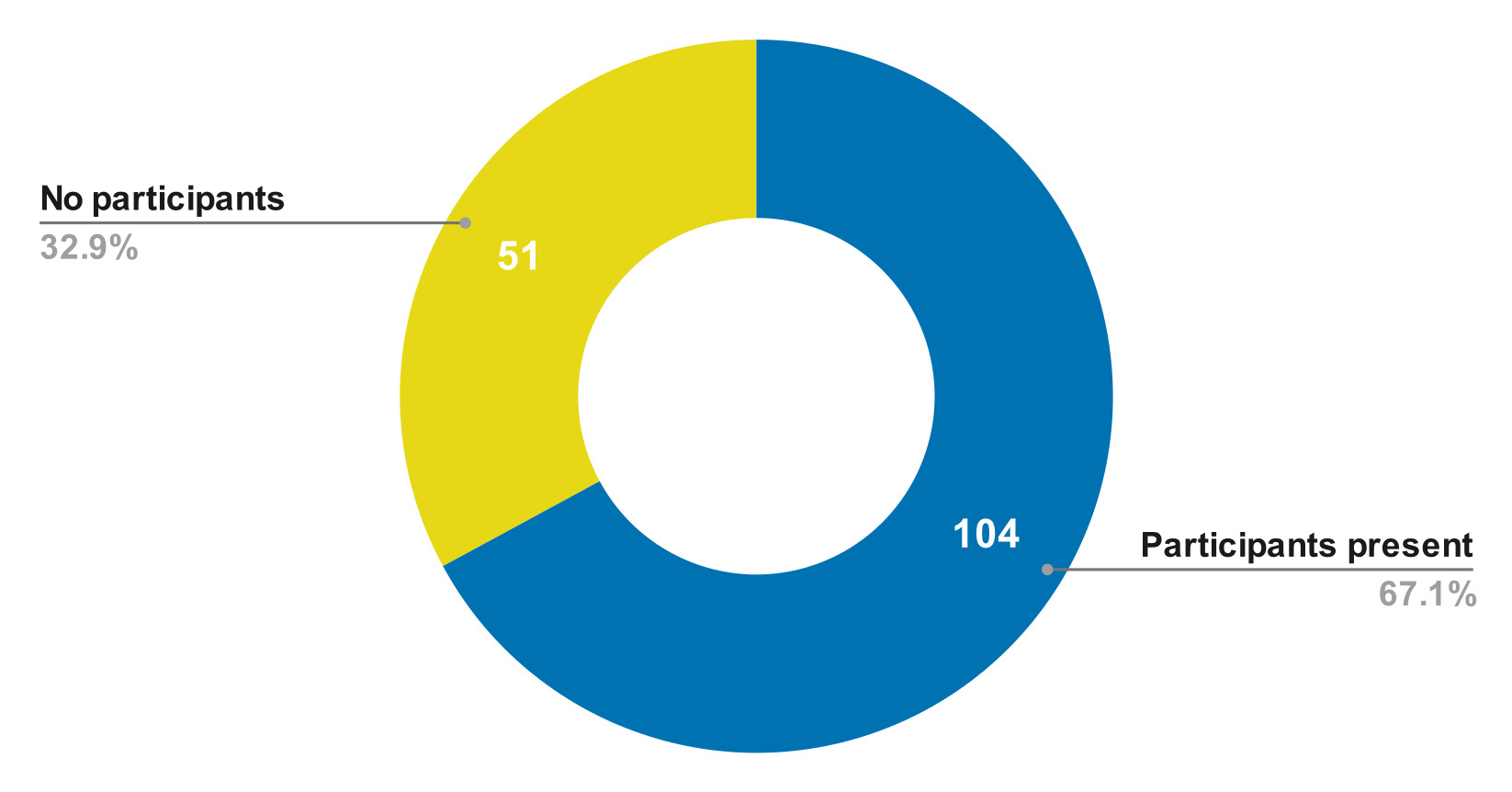}
  \caption{Distribution of usability assessment approaches depending on their requirement of actual participants. Two third of publications (67\%) involve participants in usability assessment, while one third (33\%) does not.}
  \Description{The image is a pie chart displaying the requirement of real participants in usability issue automation approaches. The larger segment of 101 publications (67\%) relies on data obtained from participants whereas a smaller segment of 51 publications (33\%) does not.}
  \label{fig:dev-p-involvement}
\end{figure}

From categorization of automation approaches based on their requirement for actual human participants (see \autoref{fig:dev-p-involvement}), it is apparent that two thirds (n=104) rely on participants to provide explicit or implicit feedback that results from genuine human cognition of, or interaction with, the assessed system. Participants mostly appear in the role of users, supplying data such as interaction logs, verbal feedback and emotions \cite{buono2020, salomon2023, villamane2024}, although they can also be experts \cite{asemi2022}. The remaining third of research (n=51) seeks to bypass the need for participant involvement altogether. 

\begin{figure}[!ht]
  \centering
  \includegraphics[width=0.6\linewidth]{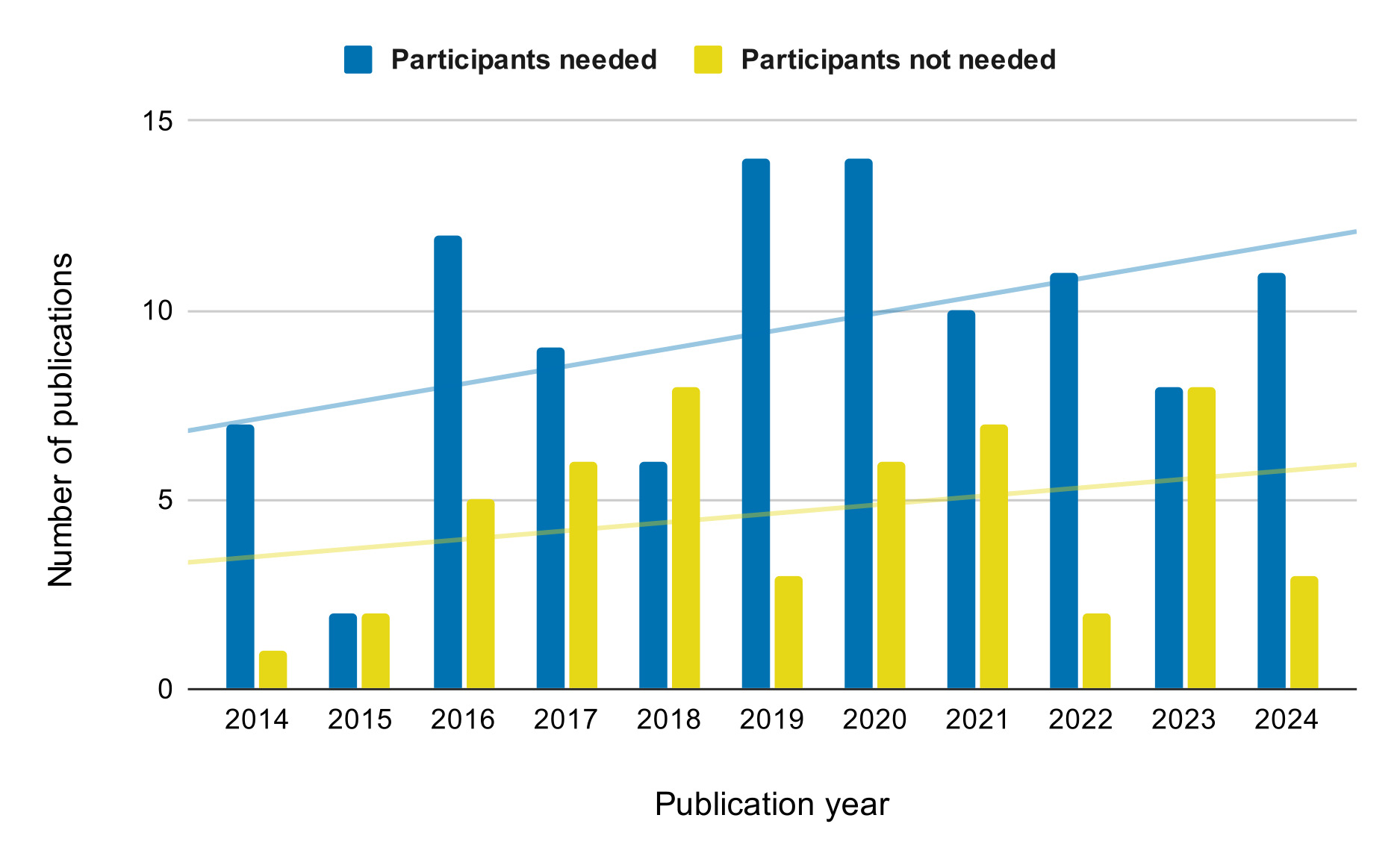}
  \caption{Evolution of participant-dependent and participant-independent usability assessment approaches in the last decade. Participant-dependent approaches have been more prevalent during most years, with the exception of 2015, 2018 and 2023.}
  \Description{The image is a vertical bar chart displaying the annual number of publications depending on the involvement of real participants for detection of usability issues (two bars for each year for with and without). The horizontal axis represents the publication year from 2014 to 2024 while the vertical axis represents the number of publications. The number of publications with real participants as well as ones without them slowly grow over time.}
  \label{fig:dev-p-involvement-years}
\end{figure}

Historically, only two years marked participant-independent approaches as equally or more prevalent than the alternative (as seen in \autoref{fig:dev-p-involvement-years}, aside from 2015 which yielded few publications on the topic overall):
\begin{itemize}
    \item 2018, when deep learning and CNNs were first applied for usability assessment based on UI images \cite{ponce2018, bakaev2018, bakaev2018a} (previous deep learning pursuits involved face recordings and physiological signals \cite{liapis2014, matlovic2016}), and
    \item 2023, when LLMs came into spotlight as having potential for generating and evaluating feedback in natural language \cite{asnawi2023, duan2023, hämäläinen2023} (only predated by a single publication in 2022 that used GPT-3 \cite{hämäläinen2022}).
\end{itemize}

It should be emphasized that there is a distinction between whether an approach itself incorporates feedback from human participants, and whether participants were involved in a study strictly for validation. Several participant-independent methods were subject to comparison with human feedback \cite{xiang2024, hämäläinen2023, hsueh2024, xing2021, cassino2015, dingli2014}. Supervised machine learning methods were also learned based on objective system-derived inputs—source code and GUI images—that were labeled by humans, thus extrapolating feedback from a different context \cite{dou2019, xing2021, hasanmansur2023}.

\section{Discussion}
\label{sec:discussion}

While various approaches exist for usability assessment, generally they can be classified into either those that draw from genuine user experiences and cognitions—usability testing and monitoring—and those that leverage the expertise of evaluators—usability inspection. However, the presentation of automated techniques sometimes blurs conceptual lines between methodological paradigms. The tip of the iceberg is embodied by studies that seek to emulate usability testing by simulating human participants \cite{xiang2024, hämäläinen2023}. In some instances, the term “usability testing” is also applied synonymously with automated usability evaluation with no actual participants \cite{paul2020, namoun2021, csontos2021}. This mindset may originate from the perception of usability testing as an extension of software testing in quality assurance \cite{marenkov2016a, robal2017, almeida2015, widodo2023}. Manually testing for bugs is ineffective, which is why it is routinely automated. Therefore, manual usability testing is also viewed as a problem in need of solving, with proposed automated solutions that seek to be less time-consuming \cite{schoop2022, zhou2020} or more objective \cite{dou2019, cassino2015, robal2017}.

However, while bugs are bottom-up problems, usability issues manifest from the collision between top-down and bottom-up factors. Human subjectivity is essential for genuine usability testing to commensurably reflect real-world complexity. Our survey revealed that when results of human-centered and fully-automated assessments are compared in detail, the significance of the human factor is evident. For example, the aesthetic perception of UIs by humans is affected by high-level design aspects, which an assessor AI could not replicate, since the features and patterns that it has learned from its training data created a bias in the absence of a capacity for greater contextual understanding \cite{dou2019}. LLM simulation of elderly and young participants interacting with an application was compared to feedback from actual usability testing \cite{xiang2024}. Humans provided nuanced insights influenced by the genuine top-down perspective of personal needs and task experiences. Although the AI feedback was phrased as if written by a person, its contents were generic and guideline-like, such as elderly persons having issues with eyesight and text readability.

The relationship between participant-free automated methods and usability testing has parallels to the traditional dichotomy between testing and inspection \cite{nielsen1994, hollingsed2007}. Just as human experts conducting usability inspection inherently introduce biases, automated approaches—whether expert systems or data-driven models–do the same. For unambiguous and representative terminology that fosters correct expectations and methodological acuity, we assert that an automated technique should only qualify as usability testing if it directly involves data from users interacting with a system in the specific context of use (e.g., not merely based on patterns learned from other systems and tasks). Referring to such methods as usability testing with the qualifier “simulated” may also be acceptable, but only if the implications are clear to the audience, considering the current discourse on AI and its societal impacts. 

To differentiate automated techniques that extrapolate extraneous usability insights from external contexts to new ones, as opposed to standard usability testing, we propose the term Usability Transpection. This neologism aligns with usability inspection in its reliance on a central repository of knowledge—whether expertise or a dataset of usability feedback from different contexts—and its similar role in the design process, while emphasizing its transitive, data-driven nature. Unlike inspection that leverages expertise directly, Usability Transpection generates results that require further expertise to interpret. The introduction of Usability Transpection is less relevant for methods that utilize genuine user feedback, but it offers a broader picture of automation for a more consistent and precise classification between Usability Testing, Inspection and Transpection.

Given that eliminating humans as a source of feedback in the name of efficiency would be an overcorrection with adverse collateral effects, emphasis should continue to be placed on increasing the efficiency of human-centered methods. The expertise of human researchers and their decision-making plays a critical role, opening the path for AI augmentation rather than full automation \cite{fan2022, esposito2024}. There are three aspects in which advanced automation (e.g., research assistants) can make collection of usability feedback from users less time-consuming and more enriched: real-time adaptation aimed at enhancing the interaction with participants \cite{kuric2024, liu2024, celino2020}, augmentative support of data analysis \cite{kuang2024, kuang2023a, batch2024} and assistance with planning and setting up of user experiments.

Low technological maturity in the field—with few tools, particularly those involving AI, validated past the prototype stage—could point to a significant research gap. The slow growth of the field is particularly noticeable in the context of the expansion of AI-driven methods, which may have potential, but require meticulous exploration and validation to be used reliably. Given the high variability of user experiences and usability issues, more ecologically valid studies along with reproduction and replication studies should also be pursued to improve the generalizability of knowledge about the applicability of techniques and instruments. 

Given the lack of reliability in some established usability evaluation tools \cite{namoun2021, parajuli2020} and the ongoing advancements in machine learning, the field is ripe for the development of innovative automation methods. To enhance theoretical understanding, the field would benefit from more in-depth analysis of the threats to validity to research methods and limitations of automated approaches. Accounting for the implications of broader classifications of methods (e.g., testing vs. inspection vs. transpection, formative vs. summative, qualitative vs. quantitative, fully-automated vs. augmentative) could facilitate their practical implementation in software development and design processes. Across the primary literature, these aspects can sometimes appear relatively streamlined \cite{dou2019, chen2021, ponce2018, blanco-gonzalo2014, kumar2023, bures2020}.

\section{Threats to validity}
\label{sec:threats}

Systematic literature reviews (SLRs) are susceptible to a multitude of threats to validity (TTVs). \citet{zhou2016} compiled a comprehensive list of TTVs in software engineering and the strategies used to address them. To ensure the high validity of our results, we adopted a number of mitigation strategies during planning, conducting and reporting stages of our SLR process (see \ref{sec:methodology} \hyperref[sec:methodology]{Methodology} for key research design decisions). The addressed threats can be broadly categorized into the following groups:

\textit{Internal validity.} To mitigate the effects of potential confounding variables, we aimed to account for biases in the selection of primary studies by retrieving them from multiple well-established and relevant library sources, using an integration of query and reference search. Researcher biases were addressed by cooperative and iterative tuning of inclusion criteria, search keywords and quality assessment. Publication bias was addressed by establishing criteria for the inclusion of preprint articles of sufficiently high quality. 

\textit{External validity.} The decision to limit the scope of the study to the last ten years to reflect up-to-date technology and knowledge may have prevented some older relevant studies from being included. Limited generalizability and ecological validity of some primary studies also restrict the strength of some claims presented in this survey.

\textit{Construct validity and Conclusion validity.} The accuracy of the coding and synthesis of studies may be affected by errors, incomplete information, and flawed presentation within the surveyed publications themselves, or due to various biases on the part of the researchers. To ensure objective evaluation of the research questions, the research protocol included multiple reviewers.

\section{Conclusion}
\label{sec:conclusion}

Automation is steadily emerging as a focal point in the research of usability assessment. With the potential to increase the effectiveness and comprehensiveness of identifying and addressing usability issues, it promises to fundamentally transform the processes by which user experiences are developed and refined. This article provides a systematic literature review, analyzing the significant developments made in the field to present a comprehensive reflection of the contemporary state of the art. As technologies like deep learning and large language models rise in prominence, there is a nascent shift of attention from heuristic approaches and summative evaluation to the gathering of more qualitative insights. For future research, expanding upon the investigation of novel techniques—along with their capabilities and limitations—will be critical. Further challenges lie in the formulation of consistent taxonomies and the creation of guidelines for the use of automated techniques, leveraging their strengths and mitigating their weaknesses. The purpose of this article is to serve as a useful resource, as well as a source of inspiration for systematic efforts aimed at overcoming barriers to user-centered design through the efficiency and scalability of automation.


\section*{Data availability statement}
\label{sec:data-statement}
Additional materials are available in our online repository: \url{https://github.com/usability-ai-research/automated-issue-detection}

\bibliographystyle{ACM-Reference-Format}
\bibliography{main}

\end{document}